\documentclass[a4paper,11pt]{article}
\pdfoutput=1 

\usepackage{jcappub,wrapfig,verbatim,amsthm,graphicx,hyperref,enumerate,enumitem} 
\newcommand\ee{\end{equation}}
\newcommand\be{\begin{equation}}
\newcommand{\bsplit}{\begin{split}}
\newcommand{\esplit}{\end{split}}
\newcommand\eea{\end{eqnarray}}
\newcommand\bea{\begin{eqnarray}}

\newcommand{\bn}{\mathbf{n}}

\newcommand{\B}{\textrm{B}}
\newcommand{\F}{\textrm{F}}
\newcommand{\bb}{\textrm{b}}
\newcommand{\f}{\textrm{f}}

\newcommand{\HH}{\mathcal{H}}
\newcommand{\LL}{\textrm{L}}
\newcommand{\M}{\textrm{M}}
\newcommand{\K}{\textrm{K}}

\newcommand{\nevol}[1]{n^{\mathrm{evol}}_{#1}}

\newcommand{\fstar}{\mathcal{F}_\star}
\newcommand{\fcut}{\mathcal{F}_c}
\newcommand{\limB}{\mathcal{F}>\mathcal{F}_c}
\newcommand{\limF}{\mathcal{F}_c>\mathcal{F}>\mathcal{F}_\star}
\newcommand{\fevol}[1]{f^{\rm evol}_{#1}}

\newcommand{\FF}{\mathcal{F}}                     
\usepackage[utf8]{inputenc}
\usepackage{subcaption}
\usepackage{booktabs}
\usepackage{cleveref}
\usepackage{placeins}
\usepackage{tikz}
\usepackage{xcolor}

\newcommand{\revedit}[1]{\textcolor{black}{#1}}

\title{\boldmath Using relativistic effects in large-scale structure to constrain astrophysical properties of galaxy populations}

\author[a]{Daniel Sobral Blanco,}
\affiliation[a]{D\'epartement de Physique Th\'eorique and Center for Astroparticle Physics, Universit\'e de Gen\`eve, Quai E. Ansermet 24, CH-1211 Geneve 4, Switzerland}

\author[a]{Camille Bonvin,}

\author[b,c]{Chris Clarkson,}
\affiliation[b]{School of Physics \& Astronomy, Queen Mary University of London, London E1 4NS, UK}
\affiliation[c]{Department of Physics \& Astronomy, University of Western Cape, Cape Town 7535, South Africa}

\author[c,d,e]{Roy Maartens}
\affiliation[d]{Institute of Cosmology \& Gravitation, University of Portsmouth, Portsmouth PO1 3FX, UK}
\affiliation[e]{National Institute for Theoretical \& Computational Sciences, Cape Town 7535, South Africa}

\emailAdd{daniel.sobralblanco@unige.ch}
\emailAdd{camille.bonvin@unige.ch}

\abstract{Upcoming large-scale structure surveys will be able to measure new features in the galaxy two point correlation function. Relativistic effects appear on large scales as subtle corrections to redshift-space distortions, showing up as a dipole and octupole when cross-correlating two different tracers of dark matter. 
The dipole and octupole are very sensitive to the evolution and magnification biases of the observed tracers which are hard to model accurately as they depend upon the derivative of the luminosity function at the flux limit of the survey. 
We show that splitting a galaxy population into bright and faint samples allows us to cross-correlate these and constrain both the evolution bias and magnification bias of the two samples -- using the relativistic odd multipoles of the correlation function, together with the even Newtonian multipoles.
Although the octupole has much lower signal-to-noise than the dipole, it significantly improves the constraints by breaking parameter degeneracies.
We illustrate this in the case of a futuristic survey with the Square Kilometre Array, and demonstrate how splitting the samples in different ways can help improve constraints. 
This method is quite general and can be used on different types of tracers to improve knowledge of their luminosity functions. Furthermore, the signal-to-noise of the dipole and octupole peaks on intermediate scales, which means that that they can deliver a clean measurement of the magnification bias and evolution bias without contamination from local primordial non-Gaussianities or from systematics on very large scales.}

\begin{document}

\maketitle

\pagebreak

\section{Introduction}\label{sec:intro}

The coming generation of large-scale structure surveys, like DESI,\footnote{\url{https://www.desi.lbl.gov}} Euclid,\footnote{\url{https://www.euclid-ec.org}} LSST\footnote{\url{https://rubinobservatory.org}} and SKA\footnote{\url{https://www.skao.int}} 
will detect billions of galaxies over a large part of the history of the Universe. These observations will provide invaluable information on the properties and dynamics of galaxies. In particular, they will allow us to determine the luminosity and redshift distribution of galaxies, which is crucial to understand how galaxies form and evolve.

Moreover, knowing the luminosity and redshift distribution of galaxies is necessary to perform some cosmological tests. For example, the large-scale structure of the Universe can be used to search for primordial non-Gaussianities in the initial fluctuations generated during inflation \cite{dePutter:2016trg,Achucarro:2022qrl}. This search is however contaminated by the presence of relativistic effects in the clustering of galaxies \cite{Bruni:2011ta,Namikawa:2011yr,Jeong:2011as,Camera:2014bwa,Camera:2014sba,Alonso:2015uua,Raccanelli:2015vla, Lorenz:2017iez,Bernal:2020pwq,Wang:2020ibf,Martinelli:2021ahc,Viljoen:2021ypp,Viljoen:2021ocx,Noorikuhani:2022bwc}, whose amplitude directly depends on the astrophysical properties of the galaxies. A precise knowledge of the luminosity and redshift evolution of galaxies is therefore essential to model relativistic effects precisely and subtract the signal from the non-Gaussian one. 

A second example is the test of the cosmological principle from the measurement of the kinematic dipole in the distribution of galaxies. If the cosmological principle is valid, the kinematic dipole measured from the temperature of the Cosmic Microwave Background (CMB) and the dipole in the distribution of astrophysical sources should be in agreement \cite{ellis1984}. Current measurements show tensions between the two dipoles \cite{Secrest:2022uvx}. In this context, measuring the redshift evolution of the astrophysical dipole could help in assessing the origin of the tension \cite{Dalang:2021ruy,Guandalin:2022tyl}. Since the amplitude of the astrophysical dipole is directly governed by the luminosity and redshift evolution of the sources \cite{Maartens:2017qoa,Nadolny:2021hti}, a precise determination of these quantities is highly relevant.

The luminosity and redshift distribution of galaxies can be directly measured by binning the population of galaxies in redshift and luminosity. In this paper, we show that these measurements can be significantly improved by measuring, in addition, the two-point cross-correlation function between two populations of galaxies. More precisely, the luminosity and redshift distribution of galaxies can be encoded into two parameters: the magnification bias $s(z)$ which denotes the slope of the cumulative number of sources above a given luminosity threshold, and the evolution bias $f^{\rm evol}$ which denotes the slope of the cumulative number of sources with redshift. These two parameters directly enter in the two-point cross-correlation function, more precisely in the dipole and octupole of this two-point function. By measuring these odd multipoles, one can constrain the magnification and evolution biases, and consequently improve our knowledge on the luminosity and redshift distribution of galaxies.

The rest of the paper is organized as follow: in Section~\ref{sec:observables} we introduce the two-point correlation function of galaxies and we show how the dipole and octupole depend on the magnification and evolution biases. These quantities are modeled in Section~\ref{sec:Modeling} for a survey like SKA. In Section~\ref{sec:analysis} we discuss the methodology to constrain the magnification and evolution biases and we present our results in Section~\ref{sec:results}.

We work in units in which the speed of light $c=1$. We consider a spatially flat perturbed Friedmann-Lemaître Universe with metric $d s^2 = a^2(\eta)[-(1+2\Psi)d \eta^2 +(1-2\Phi)\delta_{ij} d x^i d x^j]$, where $\eta$ denotes conformal time.

\section{Galaxy clustering observables}\label{sec:observables}

Galaxy surveys measure the angular position of galaxies in the sky $\bn$ and their redshift $z$. From this one can construct the galaxy number count fluctuations
\begin{equation} \label{eq:Delta}
\Delta(z,\mathbf{n})\equiv\frac{N_g(z,\mathbf{n})-\bar N_g(z)}{\bar N_g(z)}\, ,
\end{equation}
where $N_g(z,\mathbf{n})$ denotes the observed number of galaxies at redshift $z$ and direction $\bn$ and $\Bar{N}_g(z)$ is the mean number of galaxies at that redshift. At linear order in perturbation theory, this observable takes the form ~\cite{Yoo:2009au,Bonvin:2011bg,Challinor:2011bk}
\begin{equation}
       \Delta(z, \bn) = \Delta^{\mathrm{st}}(z, \bn) + \Delta^{\rm lens}(z,\bn)+ \Delta^{\rm rel}(z, \bn) +\Delta^{\rm corr}(z,\bn) \,, \label{eq:clustering_tot}
\end{equation}
where 
\begin{align}
    \Delta^{\mathrm{st}} =& b\,\delta_{m} - \frac{1}{\HH}\,\partial_r(\mathbf{V}\cdot\bn)\,, \label{Eq:clustering_std} \\
    \Delta^{\rm lens}=&\left(5s-2\right)\int_0^{r}d r' \frac{r-r'}{2rr'}\Delta_\Omega(\Phi+\Psi)\, ,\label{Eq:clustering_lens}\\
    \Delta^{\rm rel} =& \frac{1}{\HH}\partial_r\Psi + \frac{1}{\HH}\dot{\mathbf{V}}\cdot\bn+\Bigg[1-\frac{\dot{\HH}}{\HH^2}-\frac{2}{r\HH}-5\,s\Bigg(1-\frac{1}{r\HH}\Bigg) + \fevol{}\Bigg]\,\mathbf{V}\cdot\bn\,, \label{Eq:clustering_dip}\\
    \Delta^{\rm corr}=&\frac{2-5s}{r}\int_0^{r}dr'(\Phi+\Psi)+\left(3-\fevol{}\right)\mathcal{H}\nabla^{-2}(\mathbf{\nabla \cdot V})+\Psi+(5s-2)\Phi\label{Eq:clustering_corr}\\
    &+\frac{1}{\mathcal{H}}\dot{\Phi}+\left(\frac{\dot{\mathcal{H}}}{\mathcal{H}^2}+\frac{2-5s}{r\mathcal{H}}+5s-\fevol{} \right)\left[\Psi+\int_0^{r}dr'(\dot{\Phi}+\dot{\Psi})\right]\,.\nonumber
\end{align}
Here an overdot denotes derivatives with respect to conformal time, $\Delta_\Omega$ is the angular Laplacian,  $r(z)$ is the comoving distance and $\mathcal{H}(z) = H(z)/(1+z)$ is the conformal Hubble parameter. 

The dominant contribution $\Delta^{\mathrm{st}}$ in~\eqref{Eq:clustering_std} contains two terms: the first one arises from the galaxy density,  and is related to the matter density $\delta_{m}$ via the linear galaxy bias $b(z)$; the second one corresponds to the Redshift-Space Distortions (RSD) due to the peculiar velocity field of galaxies $\mathbf{V}$~\cite{Kaiser:1987qv}. The lensing contribution $\Delta^{\rm lens}$ in~\eqref{Eq:clustering_lens} is negligible for current spectroscopic surveys, but it will be relevant for the next generation of surveys at high redshift~\cite{Jelic-Cizmek:2020pkh,Euclid:2021rez,Euclid:2023qyw}. It depends on the time distortion $\Psi$ and spatial distortion $\Phi$ of the metric. The contribution $\Delta^{\rm rel}$ in~\eqref{Eq:clustering_dip} is suppressed by one power $\HH/k$ compared to the standard contribution. 
The first term in $\Delta^{\rm rel}$ is a contribution from gravitational redshift, which alters the apparent size of the redshift bins due to the time distortion affecting photons incoming from galaxies located inside potential wells. The second and third terms are Doppler contributions depending on the galaxies peculiar velocities $\mathbf{V}$ and its time derivative. Finally, the last contribution  $\Delta^{\rm corr}$ in~\eqref{Eq:clustering_corr} is suppressed by two powers $\left(\HH/k \right)^2$ compared to the standard contribution. It contains Sachs-Wolfe effects, integrated Sachs-Wolfe and time delay. 

The contributions $\Delta^{\rm rel}$ and $\Delta^{\rm corr}$ are currently negligible compared to the standard contributions $\Delta^{\mathrm{st}}$. However, the coming generation of surveys will observe very large parts of the sky, allowing us to measure the galaxy power spectrum at very large scales (i.e.\ very low $k$). In this regime, $\HH/k\sim 1$ and therefore $\Delta^{\rm rel}$ and $\Delta^{\rm corr}$ become relevant. In particular, one of the goals of the coming generation of surveys is to measure local primordial non-Gaussianities by looking at their effect on the galaxy power spectrum at small $k$, where $\Delta^{\rm rel}$ and $\Delta^{\rm corr}$ are non-negligible~\cite{Bruni:2011ta,Namikawa:2011yr,Jeong:2011as,Camera:2014bwa,Camera:2014sba,Alonso:2015uua,Raccanelli:2015vla, Lorenz:2017iez,Bernal:2020pwq,Wang:2020ibf,Martinelli:2021ahc,Viljoen:2021ypp,Viljoen:2021ocx,Noorikuhani:2022bwc}. It is therefore necessary to have a precise modelling of these two contributions, that act as a contamination to the non-Gaussian signal. From Eqs.~\eqref{Eq:clustering_dip} and~\eqref{Eq:clustering_corr}, we see that these contributions depend on the magnification bias $s(z)$ and the evolution bias $\fevol{}(z)$ of the galaxy population. To properly model the contaminations to local primordial non-Gaussianities, we need therefore a precise determination of $s$ and $\fevol{}$. In the following, we will show that we can use the dipole and octupole, generated by the $\Delta^{\rm rel}$ contribution at intermediate scales to constrain $s$ and $\fevol{}$.

The information provided by galaxy surveys is analysed using summary statistics. In this paper, we focus on the three-dimensional galaxy two-point correlation function defined as $\xi \equiv \langle \Delta(\bn, z) \Delta(\bn', z') \rangle$, including  lensing and all other relativistic effects \cite{Bertacca:2012tp,Bonvin:2013ogt,Tansella:2017rpi}. This can be expanded in multipoles of the angle formed by the line of sight and the relative orientation of the pair \cite{Bonvin:2013ogt,Bonvin:2015kuc}
\begin{equation}
\label{eq:mult}
    \xi_\ell(z,d) = \frac{2 \ell +1}{2}\,\int^{1}_{-1}\,\mathrm{d}\mu\,\xi(z, d,\mu)\,P_\ell(\mu)\,,
\end{equation}
where $\mu$ is the cosine of the angle, $d$ is the distance between the two pixels and $P_\ell(\mu)$ are the Legendre polynomials of order $\ell$. We focus on intermediate scales, i.e.\ separations $d$ between 20 and 160 Mpc$/h$. This ensures that linear perturbation theory is valid, and also that local primordial non-Gaussianities are fully negligible. In this way, we can extract $s$ and $\fevol{}$ in a robust way. For each of the multipoles of~\eqref{eq:mult}, we keep only the contribution which dominates in the range of scales that we are interested in. These multipoles depend on the population of galaxies through the galaxy bias $b$ and the magnification and evolution biases, $s$ and $\fevol{}$. Here we consider different populations of galaxies, with different luminosities, and we account for both auto-correlations and cross-correlations between different populations. This is essential to measure the dipole and octupole generated by the $\Delta^{\rm rel}$ contribution, which vanish in the case of a single population~\cite{Bonvin:2013ogt,McDonald:2009ud,Montanari:2012me,Yoo:2012se,Croft:2013taa,Paul:2022xfx,Montano:2023zhh,Jolicoeur:2024oij}. Moreover, cross-correlating different populations allow us to reduce the error from cosmic variance, improving the measurement of cosmological parameters~\cite{McDonald:2008sh}.

The standard contributions of Eq.~\eqref{Eq:clustering_std} generate three even multipoles: a monopole, a quadrupole and a hexadecapole ($\ell = 0, 2, 4$, respectively). These have been measured previously for various surveys (see e.g. the eBOSS collaboration results~\cite{BOSS:2012dmf,Anderson:2013zyy,Alam:2016hwk,Alam:2017izi,eBOSS:2020yzd}). Separating the populations into two provides 7 independent measurable signals, which are given by
\begin{align}
    \xi_{\mathrm{LK}}^{(0)}(z,d) &= \left[b_{\rm L}(z)\,b_\K(z) + \frac{1}{3}\left(b_\LL(z)+b_\K(z)\right)\,f(z) + \frac{1}{5}\,f^2(z)\right]\,\sigma^2_{8}(z)\,\mu_0(d)\,, \label{eq:mono}\\
    \xi_{\rm \mathrm{LK}}^{(2)}(z,d) &= \left[- \frac{2}{3}\left(b_\LL(z)+b_\K(z)\right) f(z) - \frac{4}{7}\,f^2(z)\right]\,\sigma_{8}^2(z)\,\mu_2(d)\,, \label{eq:quad}\\
    \xi^{(4)}(z,d) &= \frac{8}{35}\,f^2(z)\,\sigma_{8}^2(z)\,\mu_4(d)\,, \label{eq:hexa}
\end{align}
where $f=\mathrm{d}\ln(\delta_m)/\mathrm{d}\ln(a)$ is the growth rate and $\sigma_8(z)$ is the amplitude of the (linear) power spectrum on the scale of $8\; \mathrm{Mpc}/h$. The indices $\LL,\K$ label the population under consideration. In the following we will split the galaxies according to their luminosity, hence $\LL$ and $\K$ are either Bright (B) or Faint (F). The functions $\mu_\ell(d)$ are defined as
\begin{equation}
\label{eq:mu}
    \mu_\ell(d) = \frac{1}{2\pi^2}\int\,dk\,k^2\;\frac{P_{\delta\delta}(k,z=0)}{\sigma_{8}^2(z=0)}\;j_\ell(k\,d)\,, \qquad \ell=0,\,2,\,4\,.
\end{equation}
Here $P_{\delta\delta}(k,z=0)$ is the matter power spectrum at $z=0$, which is computed for any theory of gravity by solving the correspondent second order evolution equations for the matter density perturbations. In this work, we fix the cosmology to be $\Lambda$CDM and compute the power spectrum using CAMB~\cite{Lewis:1999bs, Howlett:2012mh}. In addition to their dependence on cosmological parameters, the multipoles~\eqref{eq:mono}-~\eqref{eq:quad} depend on the galaxy biases $b_{\LL,\K}(z)$. Those can be modelled for each population of galaxies, with a number of free parameters that are determined from the data, as we will describe in detail in Sec.~\ref{sec:gbias}.

The lensing contribution $\Delta^{\rm lens}$ also contributes to the monopole, quadrupole and hexadecapole, as shown in~\cite{Tansella:2017rpi,Tansella:2018sld}. This contribution is however only relevant at high redshift, above $z=1$~\cite{Jelic-Cizmek:2020pkh,Euclid:2023qyw}. In the linear regime, the dipole and octupole are only measurable at redshift below 1 and we concentrate therefore our analysis on this regime, where we can safely neglect the lensing contribution. The contributions from the two other terms, $\Delta^{\rm rel}$ and $\Delta^{\rm corr}$, to the monopole, quadrupole and hexadecapole are suppressed by at least two powers $\left(\HH/k \right)^2$~\footnote{The largest contribution from $\Delta^{\rm rel}$ is indeed through $\langle \Delta^{\rm rel} \Delta^{\rm rel}\rangle$, since $\langle \Delta^{\rm st} \Delta^{\rm rel}\rangle$ generates only odd multipoles, while the largest contribution from $\Delta^{\rm corr}$ is through $\langle \Delta^{\rm st} \Delta^{\rm corr}\rangle$.}. For the range of separations we are interested in, they are therefore negligible.

The next-to-leading order contribution originates from the cross-correlation of $\Delta^{\mathrm{st}}$ and $\Delta^{\mathrm{rel}}$, which induces a breaking of the symmetry in the correlation function, thus generating odd multipoles: a dipole and an octupole ($\ell = 1,\,3$, respectively)~\cite{Bonvin:2013ogt, McDonald:2009ud, Croft:2013taa, Yoo:2012se}. These odd multipoles are nonzero only for the cross-correlation of two different populations, $\LL=\B$ and $\K=\F$~\cite{Bonvin:2015kuc, Gaztanaga:2015jrs}. Using the Euler and Poisson equations we obtain
\begin{align}
    \xi_{\B\F}^{(1)}(z,d) &= \frac{\mathcal{H}(z)}{\mathcal{H}_0}\,\Bigg[\frac{3}{5}\left(\beta_\B(z) - \beta_\F(z)\right)f^2(z) + (b_\F(z)\,\beta_\B(z) - b_\B(z)\,\beta_\F(z))\,f(z) \nonumber \\ 
    &\qquad\qquad\qquad\qquad + (b_\B(z) - b_\F(z))\left(\frac{2}{r\mathcal{H}} + \frac{\mathcal{H}'}{\mathcal{H}^2}\right) f(z) \Bigg]\,\sigma_{8}^2(z)\,\nu_1(d)\,, \label{eq:dipole} \\
    \xi_{\B\F}^{(3)}(z,d) &= \frac{2}{5}\,\frac{\mathcal{H}}{\mathcal{H}_0}\,(\beta_\B(z) - \beta_\F(z))\,f^2(z) \,\sigma_{8}^2(z)\,\nu_3(d)\,, \label{eq:octupole}
\end{align}
where we have defined the factor
\begin{equation}
    \beta_\LL(z) = 5\,s_\LL\,\left(\frac{1}{r\mathcal{H}}-1\right) + \fevol{\LL}\,,
\end{equation}
containing the magnification and evolution biases. The integrals over the power spectrum take the form 
\begin{equation}
\label{eq:nu}
    \nu_\ell(d) = \frac{\mathcal{H}_0}{2\pi^2}\int\,dk\,k\;\frac{P_{\delta\delta}(k,z=0)}{\sigma_{8}^2(z=0)}\;j_\ell(k\,d)\,, \qquad \ell=1,\,3\, .
\end{equation}

The odd multipoles are suppressed by one factor $(\mathcal{H}/k)$ with respect to the standard contribution that generate the even multipoles, as can be seen comparing Eq.~\eqref{eq:nu} with~\eqref{eq:mu}. The signal-to-noise ratio (SNR) of the odd multipoles is therefore significantly smaller than that of the even multipoles. For this reason, the odd multipoles have not been detected yet in the linear regime~\cite{Gaztanaga:2015jrs}. Forecasts for surveys like DESI and the SKA have however shown that the odd multipoles will be robustly detected in the future, see e.g.~\cite{Gaztanaga:2015jrs,Beutler:2020evf,Saga:2021jrh,Bonvin:2023jjq,Montano:2023zhh,Jolicoeur:2024oij}. Importantly, the SNR of the odd multipoles peaks at intermediate separations, $d\in [20,160]$. It is therefore not necessary to access very large scales to measure $\Delta^{\rm rel}$. As such, the dipole and octupole provide a clean way of measuring the magnification bias and evolution bias from intermediate scales, without contamination from primodial non-Gaussianities that are relevant only at very large scales.

In the flat-sky approximation, the pure standard terms $\langle \Delta^{\rm st}\Delta^{\rm st}\rangle$ do not contribute to the odd multipoles. However, going beyond the flat-sky approximation, one finds wide-angle corrections arising from RSD that contribute to the dipole and octupole at the same order as the $\Delta^{\rm rel}$ contribution. These terms need therefore to be accounted for in the analysis. They can either be included in the modelling, as done in e.g.~\cite{Paul:2022xfx,Jolicoeur:2024oij,Castello:2024jmq}, or they can be removed from the signal by an appropriate definition of the estimators~\cite{Bonvin:2013ogt, Bonvin:2018ckp}. In any case, these terms do not impact the determination of the magnification bias and evolution bias, since they do not depend on these quantities. Here we consider the second approach and remove the wide-angle effects arising in the dipole and octupole.

The cosmological and astrophysical parameters can be constrained using galaxy clustering information. In particular, measurements of the even multipoles put constraints on the the galaxy bias $b_\LL$ of each population and on the growth rate $f$, which is particularly sensitive to modifications of gravity \cite{Gleyzes:2015rua}. The relativistic effects carry additional information through the gravitational redshift contribution that can be exploited for various applications, such as measuring the gravitational potential \cite{Sobral-Blanco:2022oel}, constraining modified gravity models~\cite{Sobral-Blanco:2021cks,Tutusaus:2022cab,Castello:2023zjr} and testing for additional interactions in the dark sector \cite{Bonvin:2018ckp,Bonvin:2020cxp,Castello:2022uuu,Bonvin:2022tii,Castello:2024jmq}. Here we adopt a different perspective: we assume that the $\Lambda$CDM model is valid, and we use the even and odd multipoles to determine the cosmological parameters, the galaxy biases $b_\LL$, the evolution biases $\fevol{\LL}$ and the magnification biases $s_\LL$.

\section{Modeling of the galaxy, magnification and evolution biases}\label{sec:Modeling}

The galaxy, magnification and evolution biases depend on the population of galaxies as well as on the characteristics of the survey. In this work we consider the futuristic Square Kilometre Array Phase 2  (SKA2) HI galaxy survey. We first review the modelling of the number of galaxies and biases derived in~\cite{Maartens:2021dqy} for this survey, and then extend it to the case of two populations. This modelling is used to determine the fiducial values of the biases, which control the amplitude of the signal. We then derive fitting functions for the different biases, and we forecast how well the parameters of the fitting functions can be determined with a survey like SKA2.

\begin{figure}[ht]
    \centering
    \includegraphics[width=0.7\textwidth]{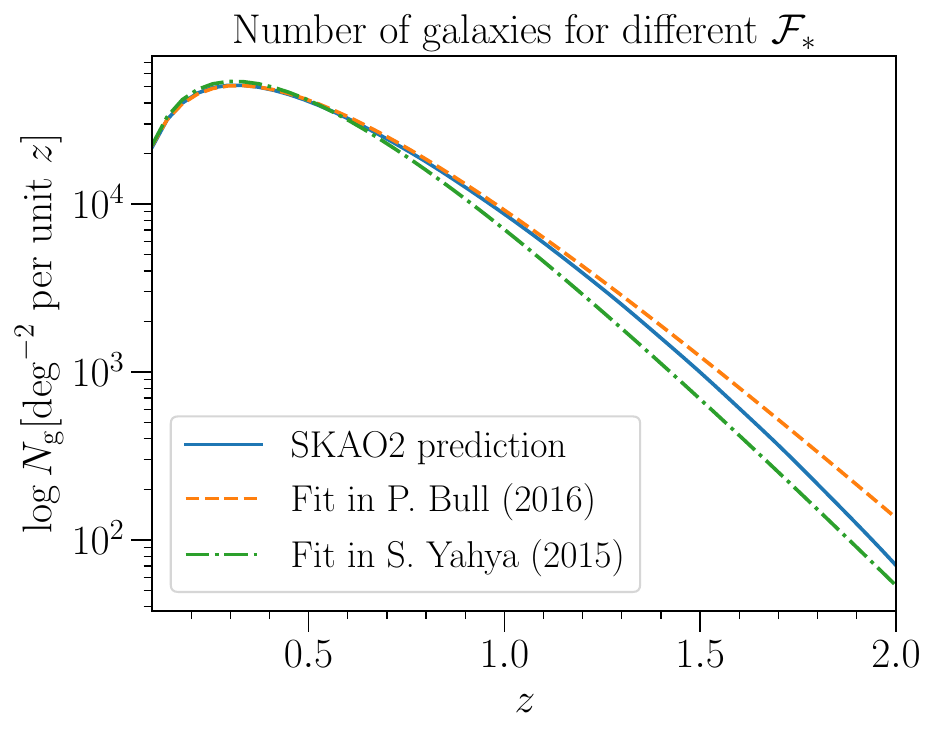}
    \caption{Logarithm of the number of galaxies per unit of solid angle and redshift for different choices of the flux limit. The solid blue line is obtained by using the $z$-dependent $\mathcal{F}_*$ from \cite{Maartens:2021dqy}. The dashed orange line is obtained by using the fit in \cite{Bull:2015lja}. Finally, the dashdot green line corresponds to the fit from \cite{Yahya:2014yva} with redshift-independent flux limit $\mathcal{F}_* = 5.0\,\mu\mathrm{Jy}$.}
    \label{fig:Ng_as_z}
\end{figure}

SKA2 is expected to detect around 1 billion HI galaxies over $30,000$ square degrees in the redshift range $z\in[0.1,\,2.0]$. Following~\cite{Maartens:2021dqy} we parametrise the background (physical) number of sources per redshift $z$ and per solid angle, detected above flux threshold $\fstar$ as
\be
\label{eq:ngfit}
\bar{N}_g(z, \fstar) = 10^{c_1}\,z^{c_2}\,e^{-c_3 z}\, ,
\ee
where the parameters $c_i$ depend on $\fstar$~\cite{Yahya:2014yva,Maartens:2021dqy}. For SKA2, the expected flux threshold will depend on redshift, as shown in~\cite{Maartens:2021dqy}, Table 2. In order to express $\bar{N}_g$ as a function of redshift alone, we first compute the $c_i(\fstar)$ using Table 3 of \cite{Yahya:2014yva}. Secondly, we reconstruct $\fstar(z)$ from Table 2 in \cite{Maartens:2021dqy}. This allows us to compute $\bar{N}_g(z)$ by replacing this $\fstar(z)$ into the interpolated $c_i(\fstar)$. We do this for the redshift interval of $z\in[0.1,\,2]$ using bins of size $\delta z = 0.1$. 
The resulting number of galaxies as a function of redshift is plotted in Fig. \ref{fig:Ng_as_z}.

\subsection{The magnification and evolution biases for a single population of galaxies}

The magnification bias $s$ is defined as the logarithmic slope of the cumulative number of sources above a given luminosity threshold. Since the luminosity threshold is related to the flux threshold by the luminosity distance $d_L$ (which does not depend on luminosity)
\be
    L_\star(z) = 4\pi\,\fstar(z)\,d_L(z)\, ,
\ee
$s$ can equally be expressed as a derivative of $\bar{N}_g(z,\fstar)$ with respect to $\fstar$ 
\begin{align}
    &s(z, L_\star)=-\frac{2}{5}\frac{\partial\log{\bar{N}_g(z, L_\star)}}{\partial\log{L_\star}}\Bigg|_z= -\frac{2}{5}\frac{\partial\log{\bar{N}_g(z,\fstar)}}{\partial\log{\fstar}}\Bigg|_z=s(z, \fstar)\,. \label{def:sbias}
\end{align}
The magnification bias is therefore an observable quantity: it can be measured by binning $\bar{N}_g(z,\fstar)$ in $\fstar$ and computing the derivative at the chosen flux threshold. Note that since the cumulative number of sources can only increase when decreasing $\fstar$, $s(z,\fstar)$ is always positive. 

The evolution bias $\fevol{}$ is defined as the partial redshift derivative of the cumulative comoving number density of galaxies above $L_\star$, $\bar{n}_g(z,L_\star)$, taken at fixed value of the luminosity threshold 
\begin{align}
\label{eq:fevol_def}
\fevol{}(z,L_\star)=-\frac{\partial \ln \bar{n}_g(z,L_\star)}{\partial \ln (1+z)}\Bigg|_{L_\star}
= -\frac{d \ln \bar{n}_g(z,L_\star(z))}{d \ln (1+z)} + \frac{\partial \ln \bar{n}_g(z,L_\star)}{\partial \ln L_\star}\, \frac{d \ln L_\star(z)}{d \ln (1+z)}
\, .   
\end{align}
The comoving number density $\bar{n}_g(z,L_\star)$ is not directly observable. It can however be related to $\bar{N}_g(z,\fstar)$ through
\begin{align}
\bar{N}_g(z,\fstar) = \frac{r^2(z)}{H(z)}\,\bar{n}_g(z,L_\star)\, ,
\label{eq:Nton}
\end{align}
where $H$ is the Hubble parameter (in physical time). Following~\cite{Maartens:2021dqy}, the partial derivative with respect to $z$ can be related to a total derivative with respect to $z$ accounting for the fact that when varying the redshift, the luminosity threshold also varies. In addition, one needs to take into account the evolution with redshift of the volume factor, $r^2(z)/H(z)$ in~\eqref{eq:Nton}, as well as the fact that the flux threshold can also vary with redshift. Putting everything together we obtain
\begin{align}
\fevol{}(z, \fstar) =& -\frac{d\ln{\bar{N}_g(z,\fstar)}}{d\ln{(1+z)}} 
        - \frac{d\ln{H(z)}}{d\ln{(1+z)}} + \frac{2(1+z)}{r(z)H(z)} \label{eq:def-fevol}\\ 
        & - 5\,s(z,\fstar) \left[1+\frac{(1+z)}{r(z)H(z)}\right]
        -\frac{5}{2}\,s(z,\fstar)\,\frac{d\ln{\fstar(z)}}{d\ln(1+z)}\,. \nonumber 
\end{align}
Written in this way, the evolution bias is an observable quantity: $\bar{N}_g$ can indeed be split in bins of redshift and the total derivative (first term) can be measured. In the following we shall refer to this quantity as $n^{\rm evol}(z)$ 
\begin{equation}
\label{eq:n_evol}
    n^{\rm evol}(z) \equiv \frac{d\log{\bar{N}_g(z,\fstar)}}{d\log{(1+z)}}\, .
\end{equation} 
The second and third term in~\eqref{eq:def-fevol} can either be obtained from measurements of $H(z)$ and $d_L(z)=(1+z)r(z)$ or from measurements of cosmological parameters in a given model. Finally, the last term in the second line is a (known) survey-dependent quantity. For surveys with a fixed flux threshold over the whole redshift range, this last term vanishes.

These results apply for the single-tracer case. Let us now extend them to the case of multiple populations of galaxies.

\subsection{Dividing galaxies into bright and faint populations}
\label{sec:splitting}

We now split galaxies into two different populations by introducing an appropriate flux cut $\fcut$. We denote galaxies with fluxes $\limB$ as \emph{bright} galaxies, and galaxies with fluxes $\limF$ as \emph{faint} galaxies
\begin{align}
\bar{N}_\B(z, \fcut)&\equiv \bar{N}_g(z, \limB)\, ,\\
\bar{N}_\F(z, \fcut)&\equiv \bar{N}_g(z, \fcut>\FF>\fstar)\, .
\end{align}
We introduce the parameter $m=\bar{N}_g(z,\fstar)/N_\B(z,\fcut)$, from which we can obtain the fraction of bright galaxies as $m^{-1}$. 
For a given value of $m$ we can find the corresponding $\fcut$ in each of the redshift bin using the fit~\eqref{eq:ngfit} to calculate both $\bar{N}_g(z,\fstar)$ and $\bar{N}_\B(z,\fcut)$ and solving numerically the equation
\be
\label{eq:m_equation}
    \log{\bar{N}_\B(z, \fcut)} - \log{\bar{N}_g(z,\fstar)} + \log{m} = 0\, .
\ee
This gives $\fcut(z)$ as a function of redshift, which is generally different from $\fstar(z)$. The number of faint galaxies, $\bar{N}_\F(z,\fcut)$, is then simply obtained by subtracting the number of bright galaxies from the total.
Note that the logarithmic evolution of the galaxies, $n^{\rm evol}$ given in~\eqref{eq:n_evol} is the same for the total population, and for the bright and faint galaxies, since $m$ is kept fixed over the redshift bins.

\subsection{Magnification bias}\label{sec:magbias}

The dipole~\eqref{eq:dipole} and octupole~\eqref{eq:octupole} depend on the magnification bias of the bright and of the faint populations. These biases multiply the Doppler contribution. They account for the fact that galaxies that are moving with respect to the observer do not have the same luminosity distance as galaxies at the same redshift with no peculiar velocities, as shown in~\cite{Bonvin:2005ps}. As a consequence their flux is magnified or de-magnified by the peculiar velocity (depending on its direction). This can change the number of bright and faint galaxies that are observed and consequently impact the correlation function.

For the bright population, galaxies can cross the flux limit $\fcut$, enhancing or decreasing the number of bright galaxies. The magnification bias of this population is therefore similar to that of the total population, simply replacing $\fstar$ by $\fcut$ in Eq.~\eqref{def:sbias}
\be
    s_\B(z) = s(z, \fcut) =  -\frac{2}{5}\frac{\partial \log{\bar{N}_g(z, \limB)}}{\partial\log{\fcut}}\Bigg|_z\, . \label{eq:sB}
\ee
The magnification bias for the faint galaxies is however more subtle: the faint galaxies can indeed either cross the $\fcut$ threshold, or the $\fstar$ threshold. Following~\cite{Bonvin:2023jjq}, we obtain the following relation for the magnification biases 
\begin{equation}
\label{eq:sFm}
    \Bar{N}_g(z,\fstar)\,s_{\rm M}(z) = \Bar{N}_\B(z,\fcut)\,s_\B(z)+\Bar{N}_\F(z,\fcut)\,s_\F(z)\,,
\end{equation}
where we have defined $s_{\rm M}(z) \equiv s(z, \fstar)$ as the magnification bias of the whole population (bright plus faint). Using that $\bar{N}_\B = \Bar{N}_g/m$, we can rewrite the above equation as 
\be
\label{eq:sF}
s_\F(z) = \frac{m}{m-1}\,s_{\rm M}(z) - \frac{1}{m-1}\,s_\B(z) \, .
\ee
Note that since by construction $\bar{N}_\B < \bar{N}_g$  we have $m>1$. 

\begin{figure}[t]
    \centering
    \includegraphics[width=0.49\textwidth]{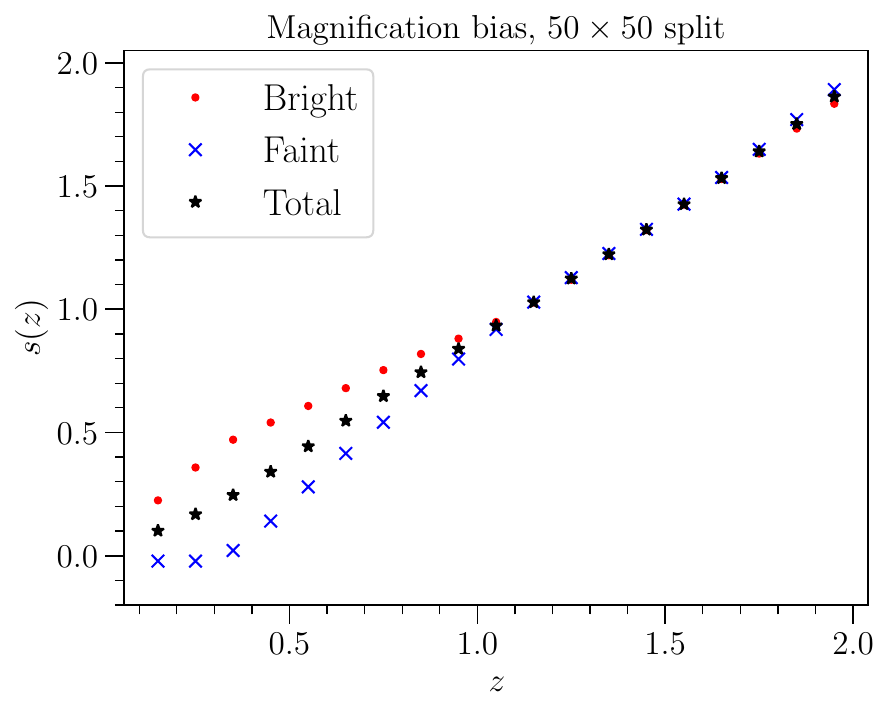}
    \includegraphics[width=0.49\textwidth]{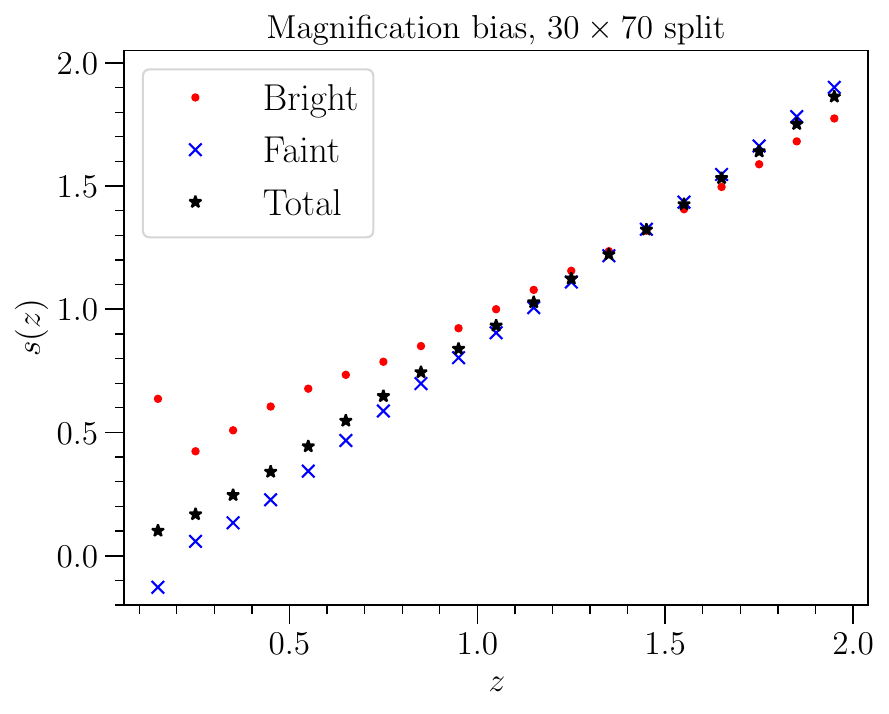}
    \caption{Predicted values of the magnification biases for the bright population $s_\B$, the faint population $s_\F$, and the total population $s_\M$. The left panel shows the magnification biases for $m=2$ (50\% of bright galaxies), and the right panel for $m=10/3$ (30\% of bright galaxies).}
    \label{fig:magbias_predictions}
\end{figure}

Contrary to $s_{\rm M}$ and $s_\B$ that are always positive, since they represent the slopes of a cumulative number of galaxies, $s_\F$ can be negative if the second term in Eq.~\eqref{eq:sF} dominates over the first one. This can happen if the slope at the flux cut, $s_\B$, is steeper than the slope at the flux threshold of the survey $s_{\rm M}$. In this case, when the fluxes are magnified by Doppler effects, more galaxies will cross the flux cut $\fcut$ (thus leaving the faint sample to become bright) than galaxies which will cross the flux threshold of the survey $\fstar$ to become detectable. In this case, a magnification of the fluxes leads to decrease in the number of faint galaxies (and vice-versa for a de-magnification). 

From Eq.~\eqref{eq:sF} we see that $s_\F$ can be directly reconstructed from $s_{\rm M}$ and $s_\B$. In our forecasts we will therefore treat $s_{\rm M}$ and $s_\B$ as free parameters, and compute $s_\F$ from these two quantities. In Fig.~\ref{fig:magbias_predictions} we show the magnification bias of the total population, and of the bright and faint populations, for two different splits: 50\% of bright galaxies (left panel) and 30\% of bright galaxies (right panel).

\subsection{Evolution bias}\label{sec:evolbias}

In addition to the dependence on magnification biases, the dipole~\eqref{eq:dipole} and octupole~\eqref{eq:octupole} depend also on the evolution bias of the bright and faint populations. The evolution bias encodes the fact that the number of galaxies is evolving with redshift, due to the formation and merging of galaxies. In a homogeneous Universe, all galaxies observed at fixed redshift $z$ are at the same distance from the observer and they share therefore the same history. However, in a inhomogeneous Universe, galaxies that are moving for example towards the observer, are at a larger distance from the observer than galaxies at the same redshift with no peculiar velocities. As a consequence, such moving galaxies live at an earlier time than the ones at rest with the Hubble flow. If the number of galaxies is evolving with time (for example decreasing due to mergers) we will see less galaxies at earlier time than closer to us, which creates fluctuations in the galaxy number density.

In the case of two populations, the evolution bias of each population encodes how each of the population evolves and it is therefore simply given by~\cite{Maartens:2021dqy,Bonvin:2023jjq}
\begin{align}
\fevol{\B}(z) &= -\frac{\partial \log{\bar{n}_\B(z,L_c)}}{\partial\log{(1+z)}}\Bigg|_{L_c}\, ,\label{eq:fevolB_def}\\
\fevol{\F}(z) &= -\frac{\partial \log{\bar{n}_\F(z,L_\star,L_c)}}{\partial\log{(1+z)}}\Bigg|_{L_\star,L_c}\, .\label{eq:fevolF_def}
\end{align}
Following the same steps as in the case of one population of galaxies, we can rewrite Eqs.~\eqref{eq:fevolB_def} and~\eqref{eq:fevolF_def} in terms of observable quantities
\begin{align}
    &\fevol{\B}(z) = - n^{\rm evol}(z) - \frac{d\log{H(z)}}{d\log{(1+z)}} + \frac{2(1+z)}{r(z)H(z)} -5\,s_\B(z)\left[1+\frac{(1+z)}{r(z)H(z)}\right] \label{eq:fevolB} \\
    &\qquad\qquad\qquad - \frac{5}{2}s_\B(z)\,\frac{d\log{\fcut}}{d\log{(1+z)}}\,, \nonumber\\
    &\fevol{\F}(z) = - n^{\rm evol}(z) - \frac{d\log{H(z)}}{d\log{(1+z)}} + \frac{2(1+z)}{r(z)H(z)} -5\,s_\F(z)\left[1+\frac{(1+z)}{r(z)H(z)}\right] \label{eq:fevolF} \\
    &\qquad\qquad\qquad - \frac{5}{2}\left[\frac{\bar{N}_g}{\bar{N}_\F}\,s_{\rm M}(z)\frac{d\log{\fstar}}{d\log{(1+z)}}-\frac{\bar{N}_\B}{\bar{N}_\F}\,s_\B(z)\frac{d\log{\fcut}}{d\log{(1+z)}}\right]\,, \nonumber
\end{align}
where $n^{\rm evol}(z)$ is given by Eq.~\eqref{eq:n_evol}. These results are consistent with Eq.~(17) of~\cite{Bonvin:2023jjq} when both $\fstar$ and $\fcut$ are constant. Note that the ratios in the second line of Eq.~\eqref{eq:fevolF} are the same as those appearing in Eq.~\eqref{eq:sF}. They are constant by construction and only depend on the fraction of bright and faint populations, encoded in $m$:
\begin{equation}
    \frac{\bar{N}_g}{\bar{N}_\F} = \frac{m}{m-1}\,,\qquad\qquad \frac{\bar{N}_\B}{\bar{N}_\F} = \frac{1}{m-1}\,.
\end{equation}

 \begin{figure}[t]
    \centering
    \includegraphics[width=0.49\textwidth]{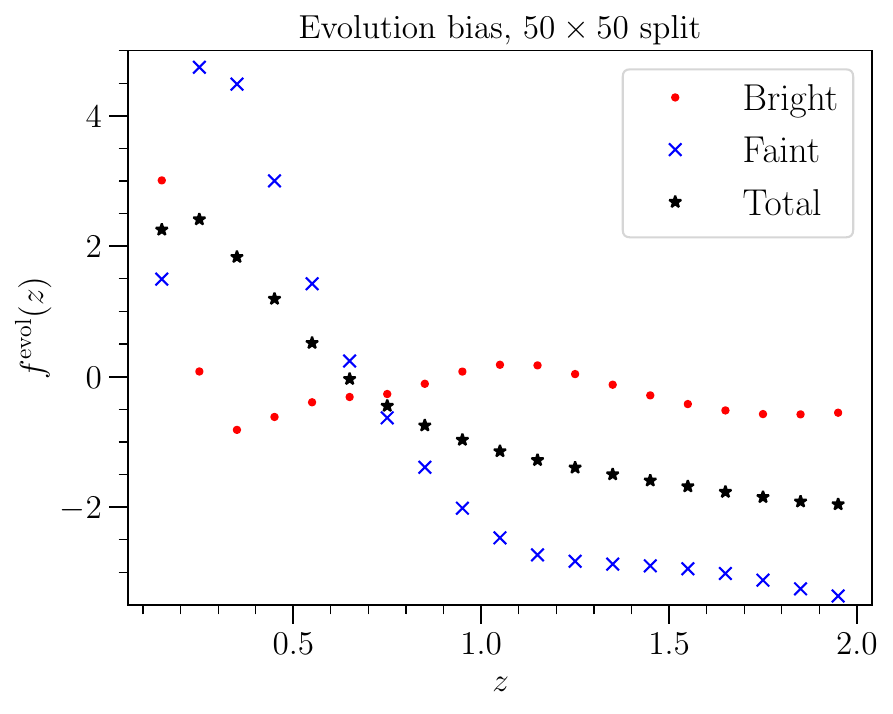}
    \includegraphics[width=0.49\textwidth]{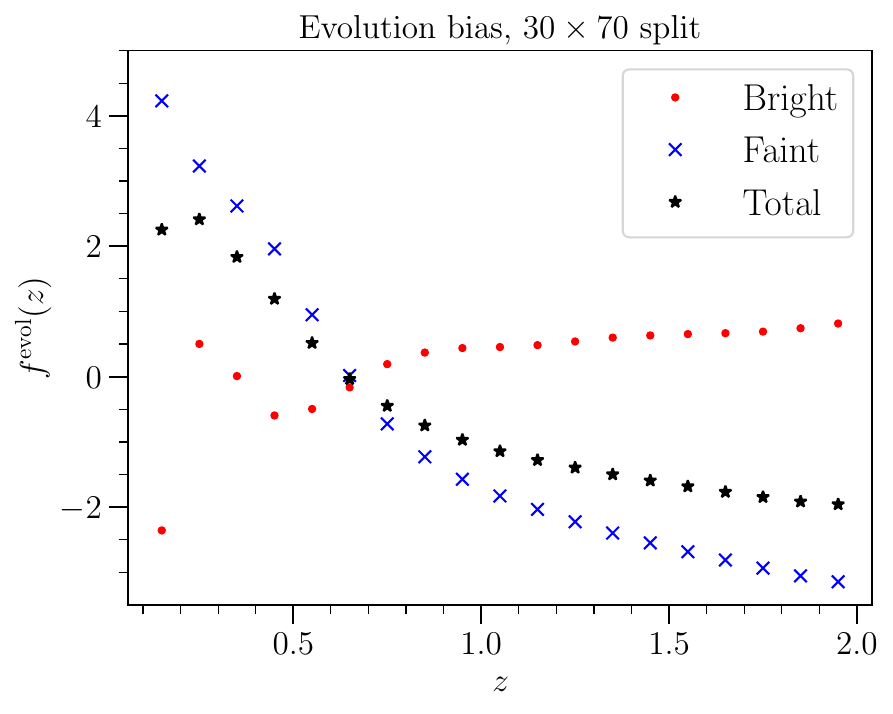}
    \caption{Predicted values of the evolution biases for the bright population $\fevol{\B}$, the faint population $\fevol{\F}$, and the total population $\fevol{}$. The left panel shows the evolution biases for $m=2$ (50\% of bright galaxies), and the right panel for $m=10/3$ (30\% of bright galaxies).}
    \label{fig:fevol_predictions}
\end{figure}

We have now all the ingredients to compute the evolution bias numerically. As shown in Sec.~\ref{sec:splitting}, the first term, $n^{\rm evol}$, in Eqs.~\eqref{eq:fevolB} and~\eqref{eq:fevolF} is the same for both populations, and equal to that of the total population. In practice, $n^{\rm evol}$ can be measured directly from the data, by binning the galaxies in redshift, and computing the total derivative. Here we perform this derivative numerically (using 5-point stencil) from the number of galaxies obtained through Eq.~\eqref{eq:ngfit}. Similarly, we compute numerically the redshift derivatives of the flux cuts. The results for the evolution bias of the whole population, and that of the bright and faint populations are shown in Fig.~\ref{fig:fevol_predictions} for two choices of splitting. We see that the amplitude of the evolution bias of the faint sample is always larger than that of the bright sample: faint galaxies therefore evolve faster than bright ones in our modelling. Note that at low redshift, the evolution biases of the two populations are quite sensitive to the choice of $m$, providing significantly different predictions.

From Eqs.~\eqref{eq:sB},~\eqref{eq:sFm}, ~\eqref{eq:fevolB} and~\eqref{eq:fevolF} we see that the magnification biases and evolution biases of the bright and faint populations depend only on three quantities: $s_\B, s_\M$ and $n^{\rm evol}$. Interestingly, two of these quantities, $s_\M$ and $n^{\rm evol}$, are the same for all splits, since they do not depend on $\fcut$. It is therefore possible to combine different splits (with different values of $m$) to improve the measurements of $s_\M$ and $n^{\rm evol}$. These two quantities fully determine the magnification bias and evolution bias of the whole population. Hence even though measuring relativistic effects requires two populations of galaxies, the quantities that can be extracted from this signal are useful to characterise the whole population. This is particularly interesting to model the amplitude of the relativistic contaminations to local primordial non-Gaussianities, as well as the amplitude of the kinematic dipole, which do not require the use of two populations.

\subsection{Galaxy bias}\label{sec:gbias}

Finally, the dipole and octupole depend also on the galaxy biases of the bright and faint populations. The bias of each population depends directly on the splitting parameter $m$, i.e.\ on the fraction of bright and faint galaxies. Typically, a larger value of $m$ (small fraction of bright galaxies) implies a larger $\fcut$, which leads to a larger bias for the bright population and therefore a larger bias difference $\Delta b$ between bright and faint sources. On the contrary a smaller value of $m$ decreases the bias difference (see e.g.~\cite{Bonvin:2023jjq} for an example in the case of the survey DESI). 

In the following we model the biases of the two populations in terms of the bias of the total population $b_{\rm T}$, the bias difference $\Delta b$ and the parameter $m$. The bright and faint biases obey the system of equations
\begin{align}
    &\frac{\bar{N}_\B}{\bar{N}_g}\,b_\B + \frac{\bar{N}_\F}{\bar{N}_g}\,b_\F = b_{\rm T}\,, \\
    &b_\B - b_\F = \Delta b\,.
\end{align}
Using  $\bar{N}_\B/\bar{N}_g = 1/m$ and $\bar{N}_\F/\bar{N}_g = (m-1)/m$ we obtain
\begin{align}
    &b_\B = b_{\rm T} + \frac{m-1}{m}\,\Delta b\,, \label{eq:gbiasB}\\
    &b_\F = b_{\rm T} - \frac{1}{m}\,\Delta b\,. \label{eq:gbiasF}
\end{align}

We model the galaxy bias of the total population following~\cite{Bull:2015lja} as
\begin{equation}
    b_{\rm T} = b_1\,e^{b_2\,z}\,,  \label{eq:gbiasT}
\end{equation}
where $[b_1,\,b_2] = [0.554, 0.783]$. Without precise knowledge of the luminosity function of the population of galaxies, we do not know the bias difference $\Delta b$ for a given splitting parameter $m$. In our forecast, we assume that for $m=2$, $\Delta b=1$, in agreement with the measurements done in the BOSS survey~\cite{Gaztanaga:2015jrs}. We then choose a smaller value of $\Delta b= 0.8$ in the case where we have $70\%$ percent of bright galaxies and a larger value of $\Delta b= 1.2$ when we have $30\%$ of bright galaxies. These choices are somewhat arbitrary, but they are only necessary because we are doing a forecast. Once we perform such an analysis on real data, the bright and faint biases will be considered as free parameters, directly measured from the even multipoles of the correlation function.

\section{Methodology}\label{sec:analysis}

We forecast the capacity of a spectroscopic survey like the SKA2 HI galaxy survey to measure the magnification and evolution biases using galaxy clustering information. To this end we first parameterise the different biases with a set of free parameters, that we then constrain with the dipole and the octupole. 

We use the Fisher formalism and compute the Fisher matrix by performing the second derivative of the natural logarithm of the likelihood with respect to the parameters of the model. Assuming Gaussianity and neglecting the dependence of the covariance on the model parameters, the Fisher matrix can be expressed as 
\begin{equation}
    \mathrm{F}_{ab} = \sum_{ij}\,\frac{\partial\,\Xi_i}{\partial\,\Theta_a}\,\left(\mathbb{C}^{-1}\right)_{ij}\,\frac{\partial\,\Xi_j}{\partial\,\Theta_b}\,,
\end{equation}
where $\Vec{\Xi}$ is the data vector, $\mathbb{C}$ is the covariance matrix and $\Theta$ is the set of parameters. We describe below how we build these two objects for the specific analysis in this work. Once the Fisher matrix is computed, we estimate the covariance matrix of the parameters as the inverse of the Fisher matrix
\begin{equation}
    \mathrm{C}_{ij} = \left(\mathrm{F}^{-1}\right)_{ij}\,.
\end{equation}
The parameter uncertainties are encoded in the diagonal of $\mathrm{C}$, while the off-diagonal terms encode the cross-correlations between the parameters. 

\subsection{Survey specifications}

\begin{figure}
    \centering
    \includegraphics[width=\textwidth]{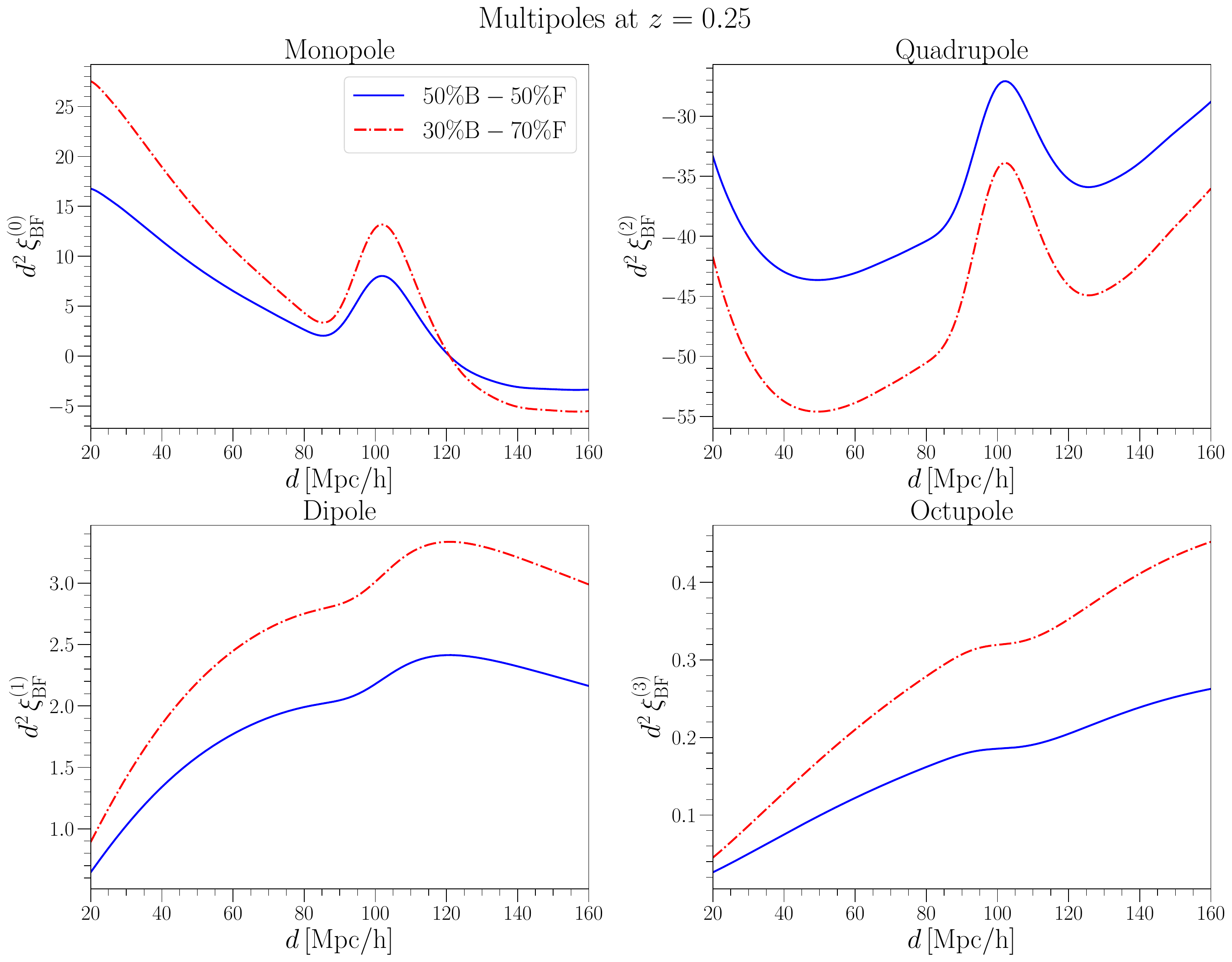}
    \caption{Multipoles of the cross-correlation between the bright and faint populations at redshift $z=0.25$, plotted as a function of separation $d$.}
    \label{fig:multipoles_z025}
\end{figure}

We consider a futuristic survey akin to the HI galaxy survey in Phase 2 of the Square Kilometer Array, as described in~\cite{Bull:2015lja,Maartens:2021dqy}. Such a survey is proposed to span $30,000$ square degrees ranging from $z=0.1$ to $z=2$, and observing close to a billion  galaxies. In practice, we divide this redshift range into $19$ bins of size $\Delta z = 0.1$ and evaluate all the quantities at the centre of each $z$-bin. We split the full population of galaxies provided by the survey into two samples following the prescription described in Sec.~\ref{sec:splitting} in order to have a constant ratio of bright versus faint galaxies in every $z$-bin. Then, we compute the signal dataset and the covariance matrices at each $z$-bin in the linear regime, including separations spanning from $d_{\rm min} = 20\, \mathrm{Mpc}/h$ to $d_{\rm max}=160\, \mathrm{Mpc}/h$. We use fiducial cosmological parameters from Planck~\cite{AghanimPLANCK2018} (see also Table~\ref{tab:fiducials_common} in Appendix \ref{app:tables}). 

In Fig.~\ref{fig:multipoles_z025}, as illustration, we show the cross-correlation monopole, dipole, quadrupole and octupole at $z=0.25$ (second redshift bin) for two different splits: 50\%  bright and 50\%  faint galaxies (blue, solid curves); 30\% bright and 70\%  faint galaxies (red, dashed-dot curves). We see that the monopole and the quadrupole are significantly larger than the dipole and octupole. The cumulative SNR over all redshifts and separations for the $50\%-50\%$ split is: $624$ and $601$ respectively for the bright-faint cross-correlation monopole and quadrupole; $57$ and $5$ respectively for the dipole and octupole.  Table~\ref{tab:SNRs_multipoles_} in Appendix~\ref{app:tables} reports the SNR for all multipoles in the different splits. Note that the SNR of the dipole and octupole are without wide-angle effects, since as explained before, one can choose an estimator to remove them.

\subsection{Fitting models for galaxy biases, magnification biases and evolution biases}\label{sec:fitting_models}

Our goal is to constrain the various biases from the data. As shown in Sec.~\ref{sec:Modeling}, the magnification and evolution biases can be written in terms of three observable quantities: $s_{\rm M}, s_\B$ and $n^{\rm evol}$, that can be directly measured from the number of galaxies. Our strategy is therefore the following: we use our modelling to compute these three quantities. We then parametrise them with a set of free parameters, with fiducial values given by our model and with uncertainties expected from future surveys. Typically, we choose two different values for the uncertainties: 50\% uncertainties and 20\% uncertainties. Those are quite large and future surveys may very well provide tighter constraints, but we prefer to remain conservative here. We then perform a Fisher forecast to determine the capability of the dipole and the octupole to improve the precision on these parameters. From this we then reconstruct $s_\B(z), s_\F(z)$, $\fevol{\B}(z)$ and $\fevol{\F}(z)$.

\subsubsection{Parametrisation of $s_{\rm M}(z)$ and $s_{\rm B}(z)$}

We parametrise the redshift evolution of $s_{\rm M}(z)$ and $s_\B(z)$ with 4 parameters each:
\begin{equation}
    \label{eq:sbias_fit}
    s_{\rm L}(z) = s_{{\rm L},0} + s_{{\rm L},1}\,z + s_{{\rm L},2}\,\log{z} + s_{{\rm L},3}\,(\log{z})^2\,,
\end{equation}
where ${\rm L}=\B,\,\M$. This expansion recovers well the redshift evolution plotted in Fig.~\ref{fig:magbias_predictions}. Thus, we include 8 parameters in the analysis: $\{s_{{\rm L},0},\, s_{{\rm L},1},\, s_{{\rm L},2},\, s_{{\rm L},3}\}$. The magnification bias of the bright population is sensitive to the choice of splitting, and we have therefore different fiducial values for the parameters for different splitting parameter $m$. On the other hand, $s_\M$ does not depend on $m$. The fiducial values can be found in Tables~\ref{tab:fiducials_sB} and~\ref{tab:fiducials_common} in Appendix~\ref{app:tables}. As explained above we introduce  Gaussian priors of $50\%$ and $20\%$ around the fiducial values on both set of parameters for each choice of $m$, to account for our prior knowledge on $s_\B$ and $s_\M$ from direct measurements.

\subsubsection{Parametrisation of $n^{\mathrm{evol}}(z)$}

Since by construction, we split the two populations keeping constant the ratio of bright and faint galaxies in all redshift bins, the function $n^{\mathrm{evol}}$ defined in~\eqref{eq:n_evol} is the same for both populations and for the total population. We parametrise the redshift evolution of this function with 4 parameters as
\begin{equation}
\label{eq:nevol_fit}
    n^{\mathrm{evol}}(z) = n^{\mathrm{evol}}_{0} + n^{\mathrm{evol}}_{1}\,z + n^{\mathrm{evol}}_{2}\,\frac{1}{z} + n^{\mathrm{evol}}_{3}\,e^{-z}\,.
\end{equation}
We have therefore 4 additional free parameters in our analysis $\{n^{\mathrm{evol}}_{0},n^{\mathrm{evol}}_{1},n^{\mathrm{evol}}_{2},n^{\mathrm{evol}}_{3}\}$. The fiducial values are independent of the value of $m$ and can be found in Table~\ref{tab:fiducials_common} in Appendix~\ref{app:tables}. Similarly to the magnification bias, we consider Gaussian priors of $50\%$ and $20\%$ around the fiducial values. 

\subsubsection{Parametrisation of the galaxy biases}

We parametrise the evolution of the biases with redshift using \eqref{eq:gbiasB}--\eqref{eq:gbiasT}. For a given splitting we fix the bias difference $\Delta b$ in Eqs.~\eqref{eq:gbiasB} and \eqref{eq:gbiasF}. The fiducial values for $b_1$ and $b_2$ are independent of the splitting since they encode the evolution of the total population. In our forecast we vary however these parameters separately when they enter into the bright or faint biases, since these two quantities can in principle vary independently. Hence we have 4 additional parameters $\{b_{1,\B},b_{2,\B}, b_{1,\F}, b_{2,\F}  \}$, with fiducial values given in Table~\ref{tab:fiducials_common} in Appendix~\ref{app:tables}. Note that we could have alternatively used as free parameters $\{b_1, b_2, \Delta b \}$.

\subsection{Galaxy samples and data vectors}\label{sec:constraints_oddmultipoles}

Since the magnification bias of the total population, $s_{\rm M}$, and the quantity $n^{\mathrm{evol}}(z)$ are independent of the way galaxies are split, we explore different cases and compare the constraints. We first perform Fisher analyses using two populations of galaxies, and then we consider a case where we have four different populations.

\subsubsection{Two galaxy populations}

We consider three different samples of bright (B) and faint (F) galaxies:
\begin{itemize}
    \item Analysis with $50\%$ of B and $50\%$ of F galaxies ($50\times 50$): we set $m=2$ and $\Delta b = 1$. This is the standard approach used in previous works, e.g.~Refs.~\cite{Castello:2022uuu, Sobral-Blanco:2022oel, Tutusaus:2022cab}. 

    \item Analysis with $30\%$ of B and $70\%$ of F galaxies ($30\times 70$): we set $m=10/3$ and $\Delta b = 1.2$. 

    \item Analysis with $70\%$ of B and $30\%$ of F galaxies ($70\times 30$): we set $m=10/7$ and $\Delta b = 0.8$.
\end{itemize}
As observables we consider the multipoles of the correlation function described in Sec.~\ref{sec:observables}. For any value of $m$, we have 7 even multipoles: the monopoles and quadrupoles of the BB, BF and FF correlations, plus the hexadecapole of the full population. Additionally, we have 2 odd multipoles arising from BF cross-correlations, namely a dipole and an octupole. The data vector at each $z$-bin takes the form 
\begin{equation}
    \Vec{\Xi}(z,d) = \left(\xi^{(0)}_{\B\B},\,\xi^{(0)}_{\B\F},\, \xi^{(0)}_{\F\F},\, \xi^{(1)}_{\B\F},\, \xi^{(2)}_{\B\B},\, \xi^{(2)}_{\B\F},\, \xi^{(2)}_{\F\F},\, \xi^{(3)}_{\B\F},\,\xi^{(4)}_\mathrm{T}\right) \!\big|_{(z,d)} \, .
\end{equation}

We compute the Gaussian covariance matrix of $\Vec{\Xi}(z,d)$ including all the possible cross-correlations between the multipoles, and accounting for both cosmic variance and shot-noise. We also include the shot-noise contributions associated to the cross-correlations between the hexadecapole and the monopoles and quadrupoles, arising from the fact that the B and F populations overlap with the full population. This contribution is usually neglected (see e.g.~\cite{Castello:2022uuu, Sobral-Blanco:2022oel, Tutusaus:2022cab}), since cosmic variance dominates the error budget of the even multipoles for a survey like the one with SKA2. In this work, we develop a shot-noise model accounting for overlapping populations (see Appendices~\ref{sec:shotnoise} and~\ref{app:hexa}) and we find that indeed the resulting contribution is negligible and has only a small impact on the constraints. 

In total we have 21 free parameters: 16 parameters describing the various biases plus 5 cosmological parameters $h, A_{\rm s}, n_{\rm s}, \Omega_{\rm b}, \Omega_{\rm m}$, whose fiducial values are also listed in Table~\ref{tab:fiducials_common} in Appendix~\ref{app:tables}.
The derivatives of the multipoles with respect to the cosmological parameters are computed numerically from CAMB~\cite{Lewis:1999bs,Challinor:2011bk} and the FFTLOG algorithm~\cite{HamiltonFFTlog} using the 5-point stencil method.\footnote{We used the Python implementation available in GitHub: \href{https://github.com/JCGoran/fftlog-python}{1D FFTlog in Python}.} The derivatives with respect to the other parameters can be computed analytically.

\subsubsection{Four galaxy populations: combining the information of two sample separations} \label{sec:constraints_joints}

We explore the possibility of further improving the constraints on the magnification and evolution biases by performing an additional Fisher analysis in which we combine two different splits. In this case, we have four populations of galaxies: two different bright populations ($\B$ and $\bb$) and two different faint populations ($\F$ and $\f$). These four populations are clearly not independent, since they are drawn from the same initial population of galaxies. In the signal, in addition to auto-correlations of each population with itself, we only consider the cross-correlations of $\B$ with $\F$ and $\bb$ with $\f$ (since parts of the galaxies in $\B$ are also in $\bb$ and similarly for the faint). The data vector is therefore $\Vec{\Xi}_{\mathrm{J}} = \left(\Vec{\Xi}_{m},\, \Vec{\Xi}_{m'}\right)$ with $m=2$ and $m'=10/3$.

In the covariance matrix we include all the cross-correlations between the multipoles of the different splits. In particular, we take into account the shot-noise contributions due to the fact that we are using the same underlying population, split in different fractions of B and F galaxies, meaning that there is certain overlap between the B and F populations of each $m$, $m'$ sub-samples. These contributions are computed in Appendix~\ref{sec:shotnoise}. Note that neglecting the shot-noise contributions between the two different splits would lead to an overestimation of the parameter constraints.

In this case we have 29 parameters: 5 cosmological parameters, 8 parameters common to the two different splits (parametrising $n^{\mathrm{evol}}$ and $s_\M$) and 16 parameters for $s_{\B}, s_{\bb}, b_{\B}, b_{\bb}, b_{\F}$ and $b_{\f}$.

\section{Results}\label{sec:results}

We first consider the cases with two populations of galaxies, and then show the constraints when combining four populations. We compare the relative reduction of uncertainties with respect to the priors on the parameters of the model when including the dipole only versus when adding the octupole, showing that in this context the octupole contains valuable information and needs to be included. We also reconstruct the $68\%$ confidence regions for the magnification and evolution biases as a function of redshift.

\subsection{Constraints from two galaxy populations}

\begin{table}[t]
    \centering
    \begin{tabular}{lrr}
        \toprule
        {} & \multicolumn{2}{c}{$50\times 50$ split} \\
        \cmidrule(lr){2-3} 
        {} &  $\xi^{(1)}_{\B\F}$ &  $\xi^{(1)}_{\B\F}+\xi^{(3)}_{\B\F}$ \\
        \midrule
        $s_{\B,0}$ &   36.81 &     30.02 \\
        $s_{\B,1}$ &   20.01 &     18.47 \\
        $s_{\B,2} $&   26.36 &     25.03 \\
        $s_{\B,3}$ &   26.18 &     23.81 \\
        \cmidrule(lr){1-3}
        $s_{\M,0}$ &   47.70 &     39.29 \\
        $s_{\M,1}$ &   31.80 &     27.83 \\
        $s_{\M,2}$ &   49.26 &     45.07 \\
        $s_{\M,3}$ &   49.99 &     49.94 \\
        \midrule
        $\nevol{0}$ &   44.85 &     41.31  \\
        $\nevol{1}$ &   43.74 &     31.04  \\
        $\nevol{2}$ &   40.39 &     25.55  \\
        $\nevol{3}$ &   33.16 &     30.71  \\
        \bottomrule
    \end{tabular} 
    \begin{tabular}{rr}
        \toprule
        \multicolumn{2}{c}{$30\times 70$ split} \\
        \cmidrule(lr){1-2}
        $\xi^{(1)}_{\B\F}$ &  $\xi^{(1)}_{\B\F}+\xi^{(3)}_{\B\F}$  \\
        \midrule
        41.58 &     32.77  \\
        39.44 &     36.00  \\
        44.66 &     43.48  \\
        24.08 &     21.43  \\
        \cmidrule(lr){1-2}
        47.97 &     37.71  \\
        20.88 &     18.49  \\
        49.53 &     46.46  \\
        49.98 &     49.96  \\
        \midrule
        41.32 & 40.50 \\
        37.21 & 28.01 \\
        33.77 & 24.09 \\
        31.39 & 29.61 \\
        \bottomrule
    \end{tabular}
    \begin{tabular}{rrr}
        \toprule
        \multicolumn{2}{c}{$70\times 30$ split} \\
        \cmidrule(lr){1-2}
        $\xi^{(1)}_{\B\F}$ &  $\xi^{(1)}_{\B\F}+\xi^{(3)}_{\B\F}$ \\
        \midrule
        26.65 &     23.28 \\
        17.44 &     16.73 \\
        24.12 &     22.87 \\
        28.86 &     26.80 \\
        \cmidrule(lr){1-2}
        46.19 &     37.30 \\
        28.56 &     26.66 \\
        48.01 &     41.23 \\
        49.94 &     49.84 \\
        \midrule
        46.17 &     41.65 \\
        47.18 &     32.74 \\
        32.34 &     26.57 \\
        35.96 &     31.62 \\
        \bottomrule
    \end{tabular}
    \caption{Relative 1$\sigma$ uncertainties for the magnification biases and number evolution parameters, assuming $50\%$ Gaussian priors. We show the results for 3 different splits, including only the dipole ($\xi^{(1)}_{\B\F}$) and including both the dipole and octupole ($\xi^{(1)}_{\B\F}+\xi^{(3)}_{\B\F}$). In all cases the even multipoles (monopole, quadrupole and hexadecapole) are included.}
    \label{tab:constraints_magbias_prior50}
\end{table}

\begin{figure}
    \centering
    \includegraphics[width=0.49\textwidth]{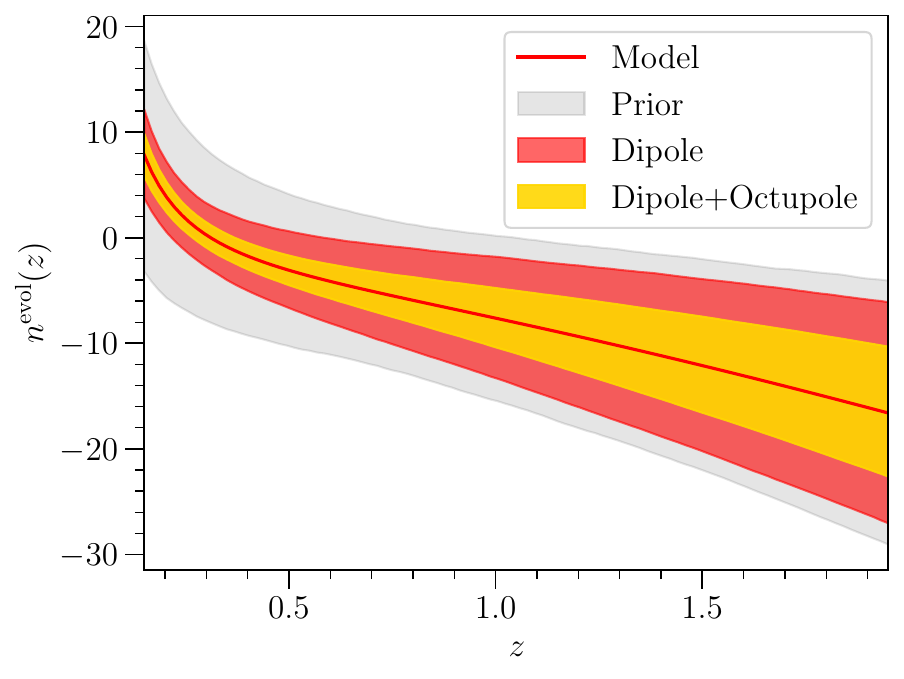}
    \includegraphics[width=0.48\textwidth]{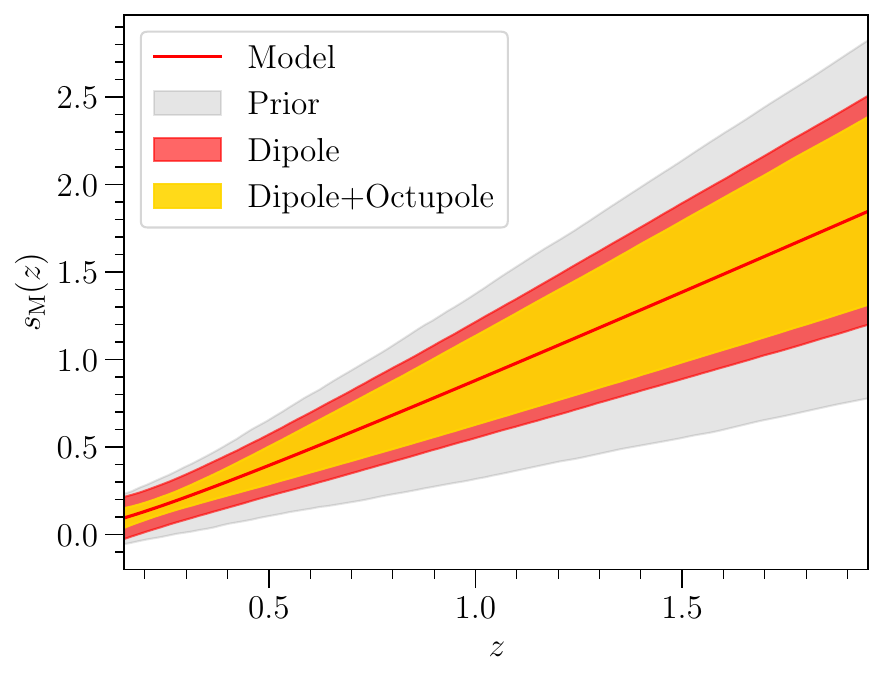}
    \includegraphics[width=0.49\textwidth]{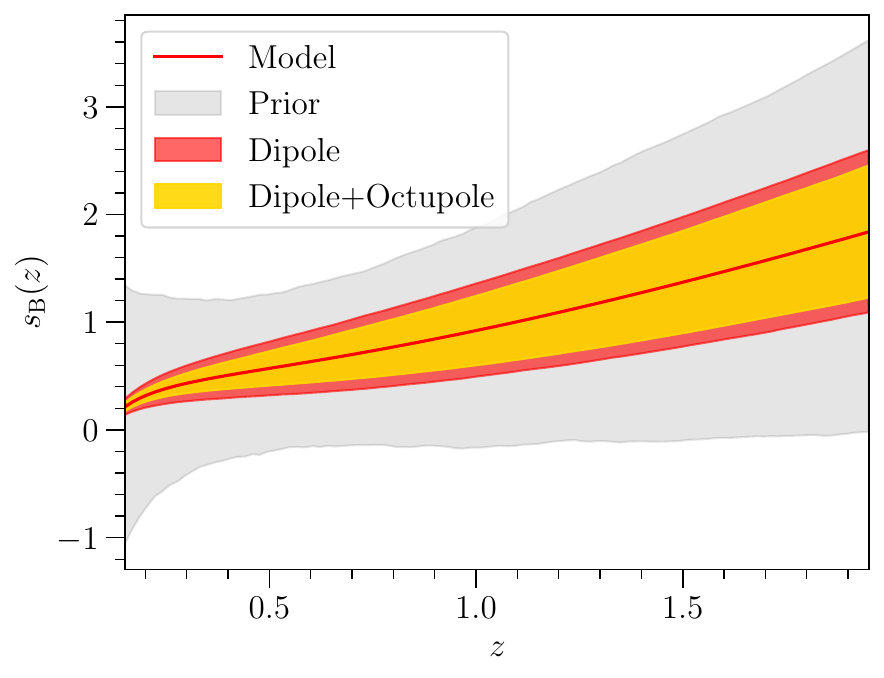}
    \includegraphics[width=0.49\textwidth]{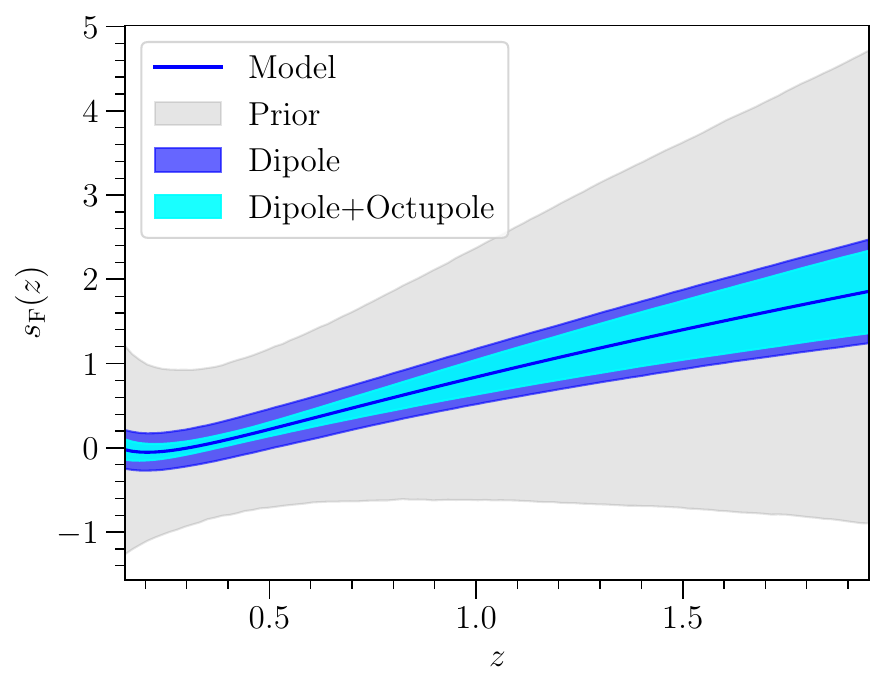}
    \includegraphics[width=0.49\textwidth]{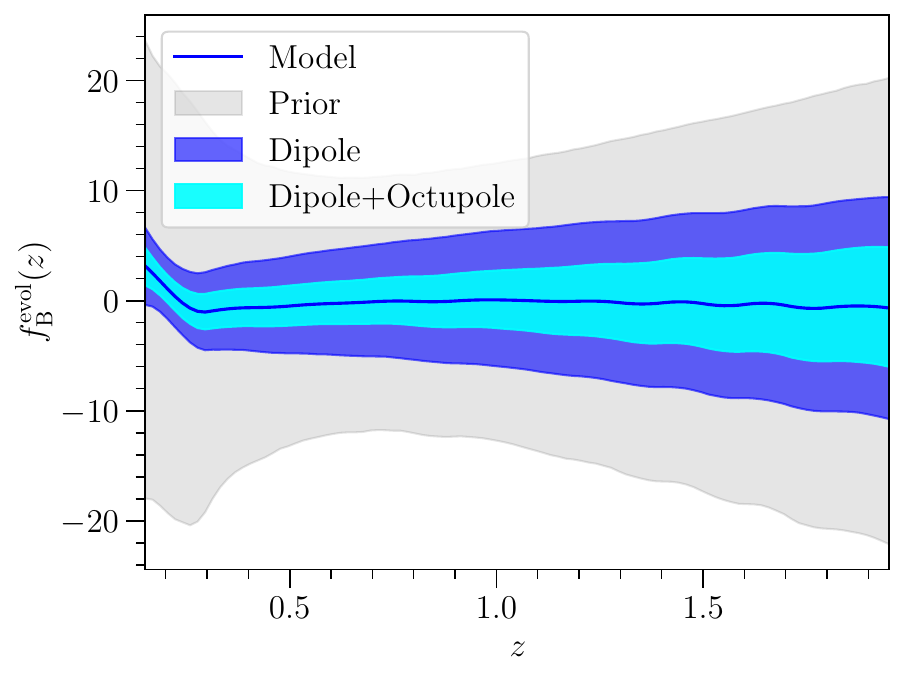}
    \includegraphics[width=0.49\textwidth]{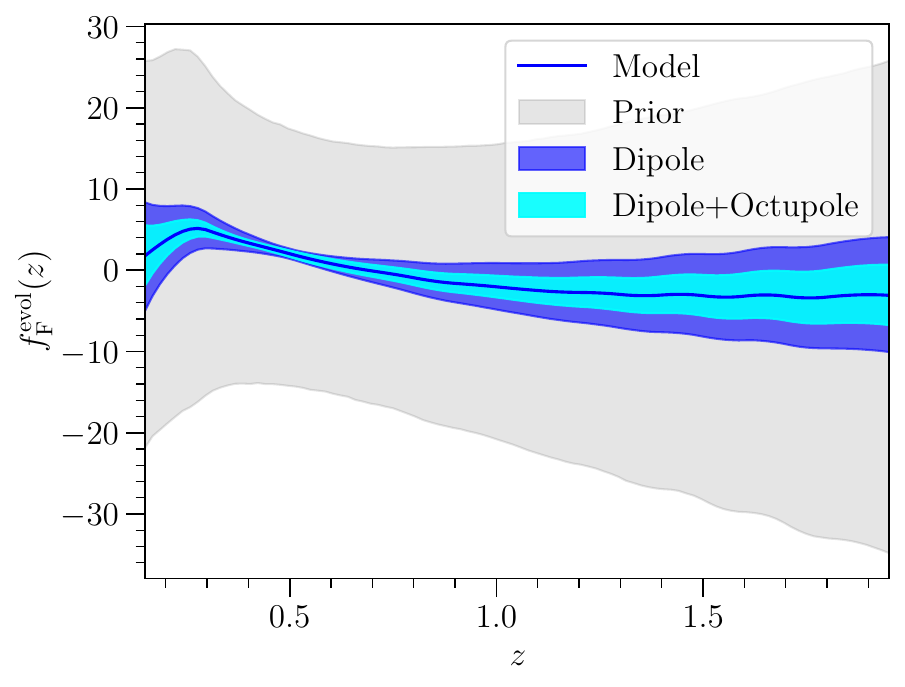}
    \caption{Constraints on the redshift evolution of the various functions for the 50$\times$50 split, starting with a 50\% prior on the parameters (grey region). We show the results when only the dipole is used and when both the dipole and octupole are included. In all cases the even multipoles (monopole, quadrupole and hexadecapole) are included. The red functions are those entering the forecasts, while the blue ones are derived from the red.}
    \label{fig:split50_prior50}
\end{figure}

In Table~\ref{tab:constraints_magbias_prior50} we show the relative 1$\sigma$ errors for the 16 parameters entering the magnification and evolution biases, in the case with 50\% priors. We show the relative errors for the three different splits. We see that the odd multipoles provide constraints on all parameters, reducing the uncertainty with respect to the priors. The improvement with respect to the prior differs significantly depending on the parameters, ranging from a reduction of the uncertainty of less than a percent (for $s_{\rm M,3}$) to a reduction of a factor $2-3$ (for $s_{\B,1}, s_{\B,2}$ and $s_{\B,3}$ in the $50\times50$ and $70\times30$ splits).

We show the relative uncertainty when only the dipole is included, and when both the dipole and the octupole are used. We see that even though the SNR of the octupole is 10 times smaller than that of the dipole, adding the octupole in the analysis has a significant impact on some of the parameters. For example, adding the octupole improves the constraints on $n_1^{\rm evol}$ and $n_2^{\rm evol}$ by 20-30\% with respect to the constraints from the dipole only, for all the splits. From Eqs.~\eqref{eq:dipole} and~\eqref{eq:octupole}, combined with the expressions for the evolution biases~\eqref{eq:fevolB} and~\eqref{eq:fevolF}, we see that the dipole depends on the magnification bias difference $s_\B-s_\F$, on the combination $b_\F s_\B-b_\B s_\F$, as well as on the product $(b_\B-b_\F)n^{\rm evol}$. On the contrary, the octupole depends only on the magnification bias difference $s_\B-s_\F$. As a consequence, there is a strong degeneracy between $n^{\rm evol}$, $s_\B$ and $s_{\rm M}$ when only the dipole is included. Adding the octupole, which does not depend on $n^{\rm evol}$ partially lifts this degeneracy, improving the constraints on the parameters. Moreover, the redshift evolution of the dipole and the octupole differ. The octupole is proportional to $f^2(z)(1-1/r(z)\HH(z))$, whereas the dipole contains also other terms that scale with $f(z)$ and with the biases $b_\B(z)$ and $b_\F(z)$. As a consequence, adding the octupole breaks degeneracies between the different parameters governing the redshift evolution of $s_\B, s_{\rm M}$ and $n^{\rm evol}$. Hence even though we find that the diagonal Fisher elements for the dipole are always significantly larger than for the octupole (by a factor 25-80 depending on the parameters), the octupole significantly helps to constrain the parameters by breaking degeneracies.

We find similar results when the priors are reduced to 20\% (see Table~\ref{tab:constraints_magbias_prior20} in Appendix~\ref{app:tables}). In this case the uncertainty on $s_{\B,1}, s_{\B,2}, s_{\B,3}$ are reduced by a factor $\sim 2$, for the $50\times50$ and $70\times30$ splits. Generally, we see that the improvement with respect to the priors are similar for the 3 splits we have chosen. Some parameters are better constrained in one split than in another, but there is no systematic trend that would single out one of the splits.

In Fig.~\ref{fig:split50_prior50}, we show the constraints on the redshift evolution of the various functions for the $50\times50$ split: the functions entering our forecasts ($n^{\rm evol}, s_\M$ and $s_\B$ in red), and the derived ones ($s_\F, \fevol{\B}$ and $\fevol{\F}$ in blue). We show the prior in grey (50\% in this case), as well as the uncertainty when only the dipole is included, and when both the dipole and octupole are included. We see that the improvement on the redshift evolution is significant for all functions. Adding the octupole helps constraining all parameters, and the improvement is especially relevant for $n^{\rm evol}$ and the two evolution biases. Similar results are found using a 20\% prior instead of a 50\% prior (see Fig.~\ref{fig:split50_prior20} in Appendix~\ref{app:tables}) and for the other splits (see Figs.~\ref{fig:split30_prior50} and \ref{fig:error_widths} in Appendix~\ref{app:tables}).

\subsection{Constraints from the combined analysis}\label{sec:results_joint}

\begin{table}[t]
    \centering
    \begin{tabular}{lrrrrrr}
        \toprule
        {} &  $50\times 50$ &  Joint  &  $\Delta_{m=2}(\%)$ &  $30\times 70$ &  Joint  &  $\Delta_{m=10/3}(\%)$ \\
        \midrule
        $s_{\B,0}$ &  30.02 &     23.72 &        \textbf{20.96} &  32.77 &     30.89 &         \textbf{5.74} \\
        $s_{\B,1}$ &  18.47 &     13.35 &        \textbf{27.73} &  36.00 &     33.36 &         \textbf{7.33} \\
        $s_{\B,2}$ &  25.03 &     19.53 &        \textbf{21.97} &  43.48 &     42.91 &         \textbf{1.33} \\
        $s_{\B,3}$ &  23.81 &     18.75 &        \textbf{21.22} &  21.43 &     21.18 &         \textbf{1.19} \\
        \cmidrule(lr){1-7}
        $s_{\M,0}$ &  39.29 &     33.61 &        \textbf{14.45} &  37.71 &     33.61 &        \textbf{10.87} \\
        $s_{\M,1}$ &  27.83 &     16.42 &        \textbf{40.99} &  18.49 &     16.42 &        \textbf{11.20} \\
        $s_{\M,2}$ &  45.07 &     40.30 &        \textbf{10.58} &  46.46 &     40.30 &        \textbf{13.25} \\
        $s_{\M,3}$ &  49.94 &     49.82 &        \textbf{0.26} &  49.96 &     49.82 &         \textbf{0.28} \\
        \cmidrule(lr){1-7}
        $\nevol{0}$  &  41.31 &     39.95 &        \textbf{3.29} &  40.50 &     39.95 &         \textbf{1.36} \\
        $\nevol{1}$  &  31.04 &     26.66 &        \textbf{14.13} &  28.01 &     26.66 &         \textbf{4.82} \\
        $\nevol{2}$  &  25.55 &     19.57 &        \textbf{23.40} &  24.09 &     19.57 &        \textbf{18.76} \\
        $\nevol{3}$  &  30.71 &     27.32 &        \textbf{11.03} &  29.61 &     27.32 &         \textbf{7.73} \\
        \bottomrule
    \end{tabular}
    \caption{Relative 1$\sigma$ uncertainties for the magnification biases and the number evolution, assuming $50\%$ Gaussian priors and including all multipoles. We show the results for a single split and for the joint analysis. The bold numbers are the relative improvement with respect to the results from a single split. Note that since the $s_{\B,i}$ parameters differ for the two splits, we have two sets of parameters in the joint analysis. Hence we show the constraints for the $50\times 50$ $s_{\B,i}$ parameters in column 2 and 3, and the constraints for the $30\times 70$ $s_{\B,i}$ parameters in column 5 and 6.}
    \label{tab:joint_prior50}
\end{table}

\begin{table}[th]
    \centering
    \begin{tabular}{lrrrrr}
        \toprule
        {} &  $50\times 50$ &  $30\times 70$ &  Joint &  $\Delta_{m=2}(\%)$ &  $\Delta_{m=10/3}(\%)$ \\
        \midrule
        $h$    &  1.1045 &   1.0767 &   1.0536 &        \textbf{4.61} &        \textbf{2.15} \\
        $\ln{\left(10^{10}\,A_{\rm s}\right)}$   &  0.7080 &   0.6890 &   0.6710 &        \textbf{5.23} &        \textbf{2.61} \\
        $n_{\rm s}$  &  0.9005 &   0.8715 &   0.8449 &        \textbf{6.18} &        \textbf{3.05} \\
        $\Omega_{\rm b}$  &  1.1296 &   1.1103 &   1.0928 &        \textbf{3.26} &        \textbf{1.58} \\
        $\Omega_{\rm m}$  &  0.4963 &   0.4793 &   0.4662 &        \textbf{6.05} &        \textbf{2.74} \\
        \bottomrule
    \end{tabular}
    \caption{Relative 1$\sigma$ uncertainties for the cosmic parameters. We show the results for two individual splits and for the joint analysis. The bold numbers show the relative improvement with respect to a single split.
    }
    \label{tab:joint_cosmic}
\end{table}

In Table~\ref{tab:joint_prior50} we show the  constraints obtained when combining the $50\times50$ split with the $30\times70$ split and we compare them with the constraints from a single split. The improvement is non-negligible for most parameters, and can even reach $20-40\%$. Hence even if adding a new split adds 4 new parameters to the Fisher (due to the magnification bias, which differs in different splits), the common parameters $s_{\M,i}$ and $n^{\rm evol}_i$ are improved by the additional signals. Interestingly, this improvement in the common parameters also impacts the individual $s_{\B,i}$ that are better measured due to the breaking of degeneracies brought by the additional split.

\begin{figure}
    \centering
    \includegraphics[width=\textwidth]{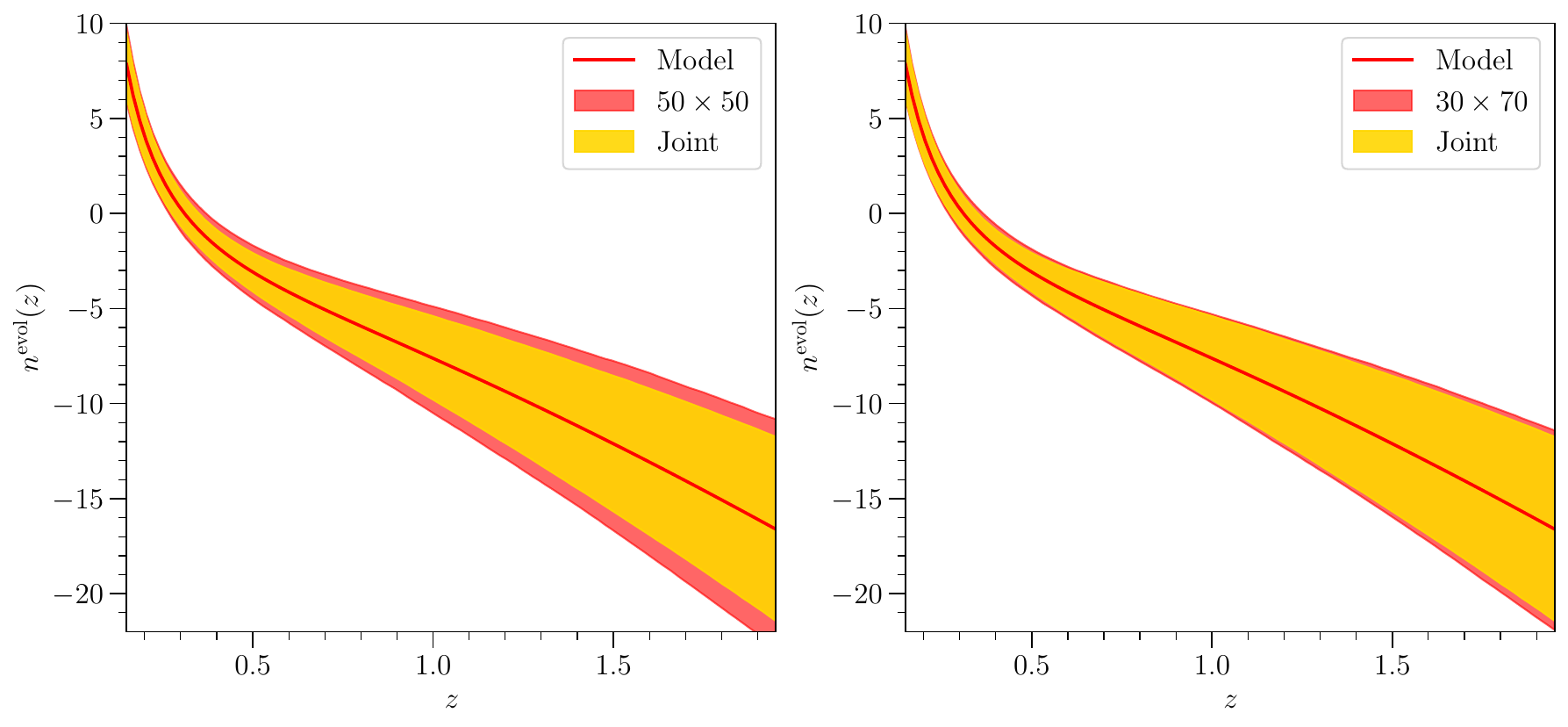}
    \includegraphics[width=\textwidth]{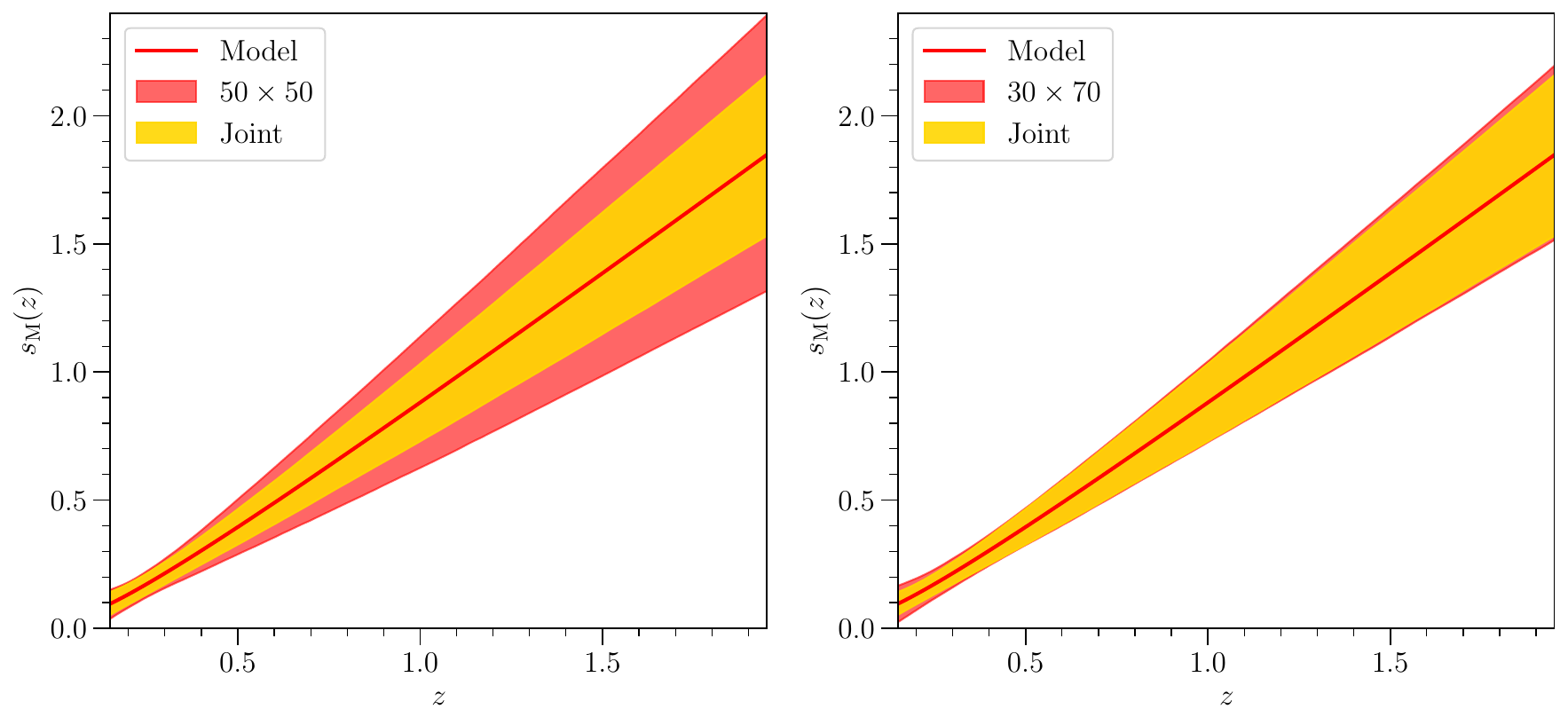}
    \includegraphics[width=\textwidth]{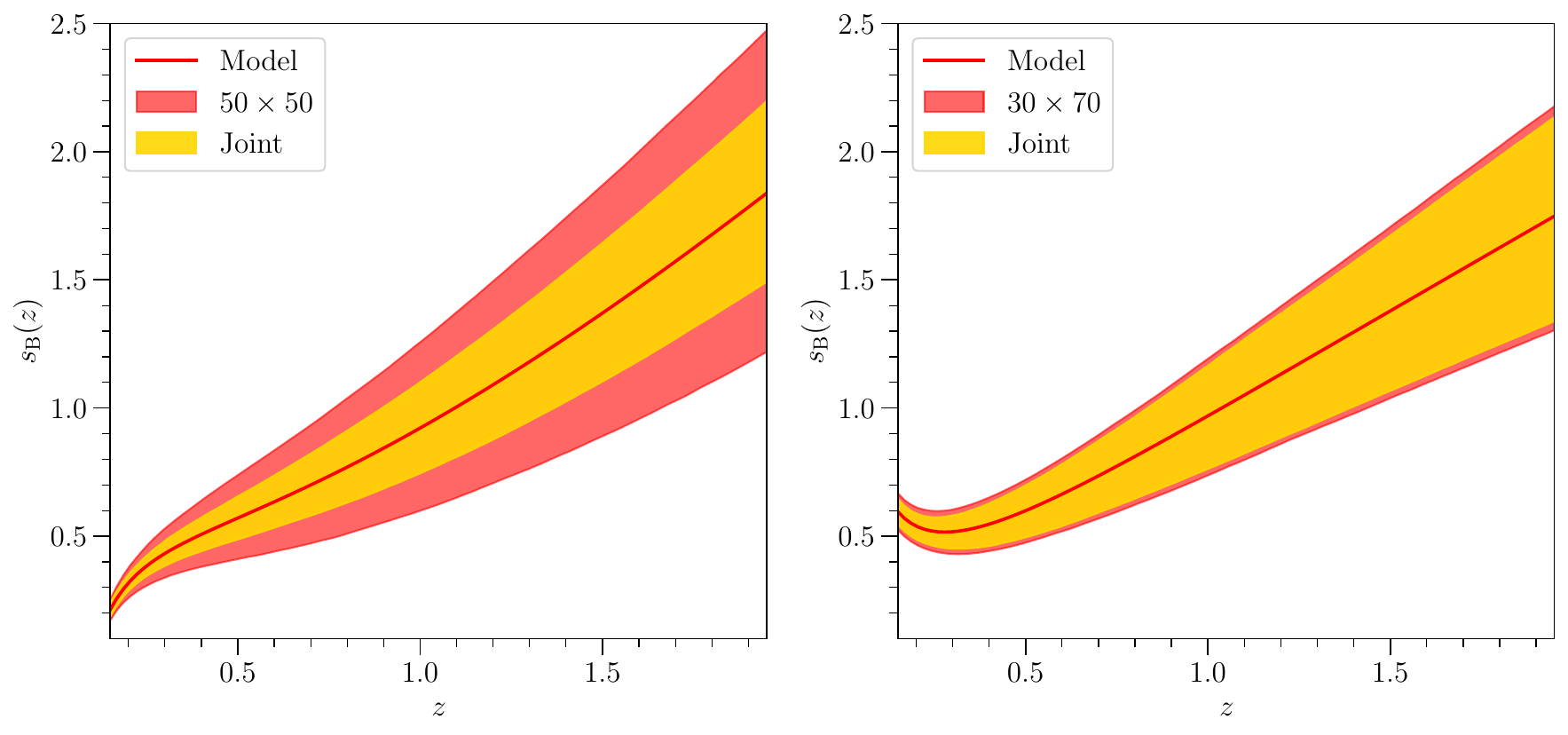}
    \caption{
    Constraints on the redshift evolution of the functions entering the forecasts, assuming a 50\% prior on the parameters. We show the results for a single split ($50\times 50$ on the left, $30\times 70$ on the right) and for the joint analysis. Note that $s_\B$ is different in the two splits. In all cases all multipoles are included.}
    \label{fig:joint}
\end{figure}

\begin{figure}
    \centering
    \includegraphics[width=\textwidth]{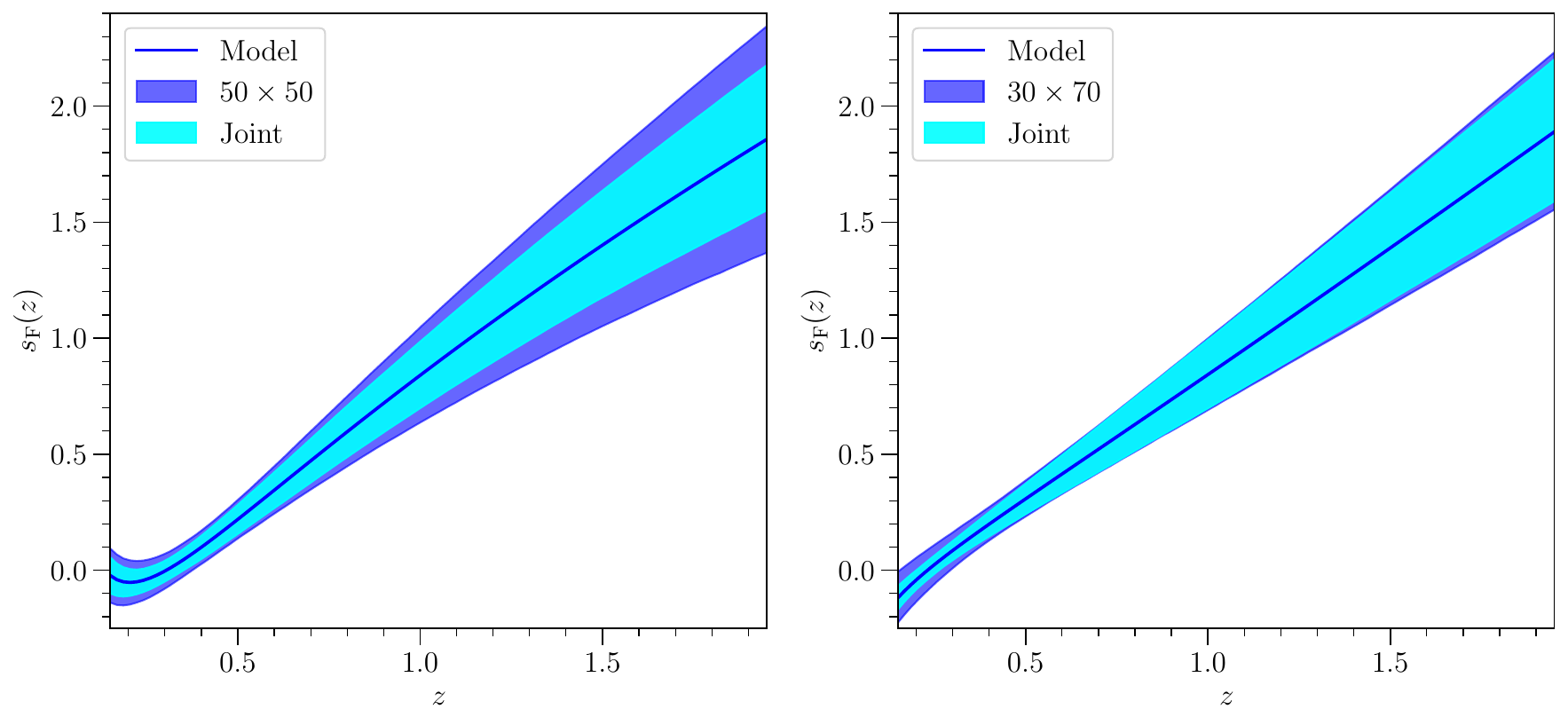}
    \includegraphics[width=\textwidth]{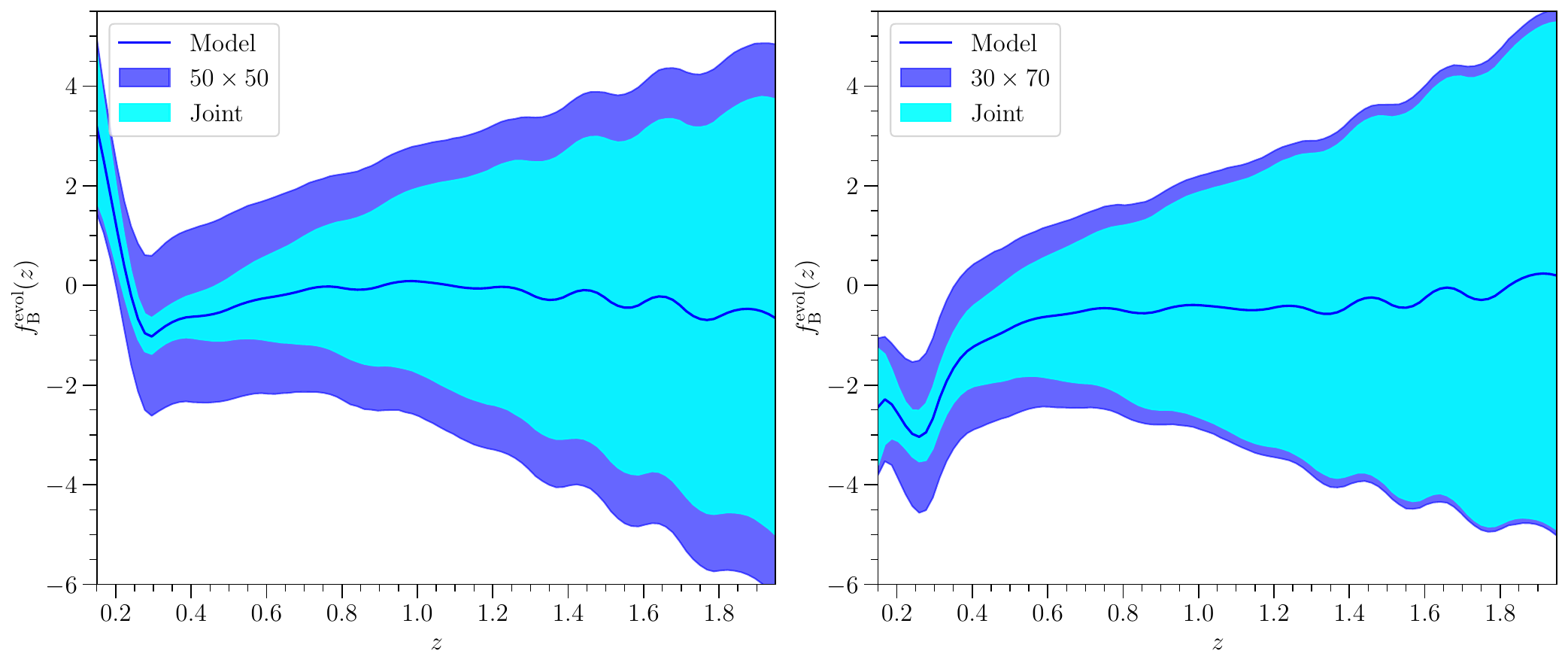}
    \includegraphics[width=\textwidth]{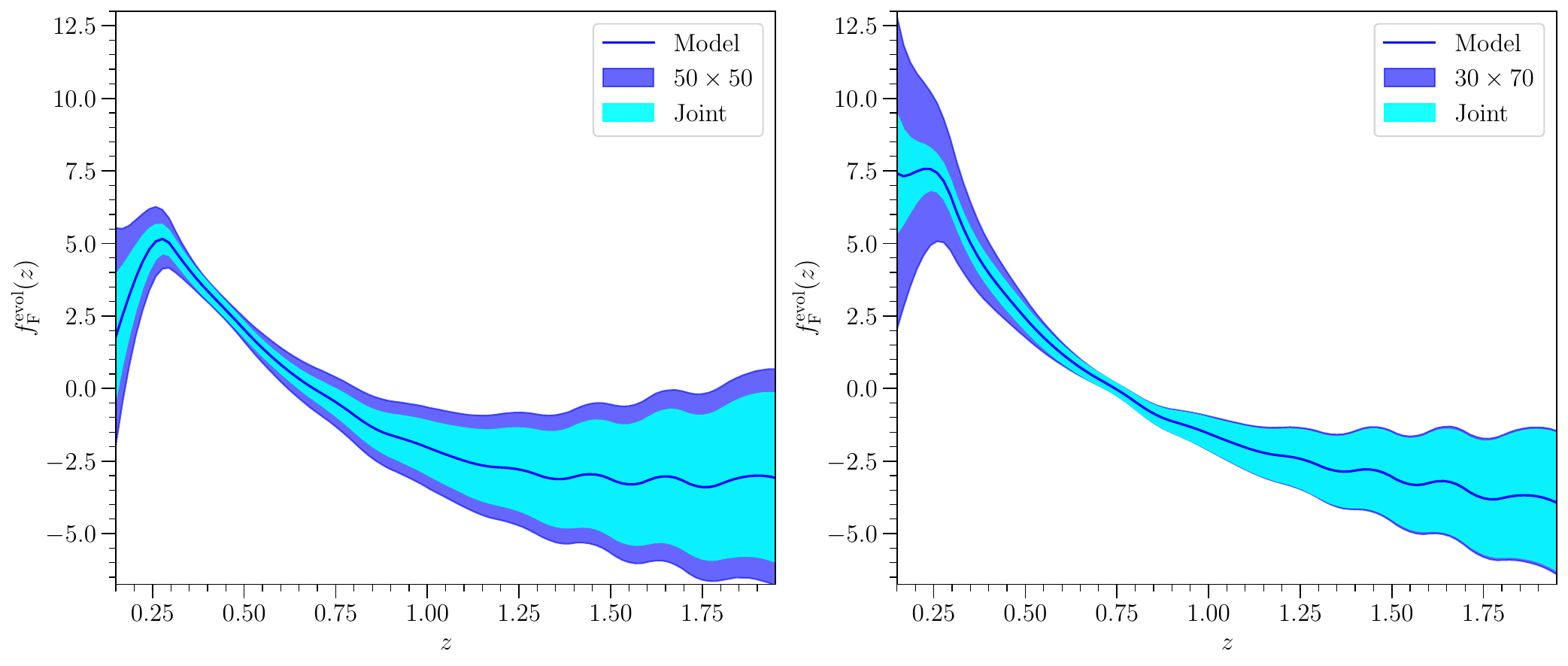}
    \caption{Constraints on the redshift evolution of the derived functions, assuming a 50\% prior on the parameters. We show the results for a single split ($50\times 50$ on the left, $30\times 70$ on the right) and for the joint analysis. Note that the three functions are different in the two splits. In all cases all multipoles are included.}
    \label{fig:jointbis}
\end{figure}

In Fig.~\ref{fig:joint} we show the constraints on the redshift evolution of the functions entering the forecasts and in Fig.~\ref{fig:jointbis} on the derived functions. In general, we see that the improvement is less significant in the case of the $30\times70$ split, but they are still non-negligible, especially for the evolution biases. Note that the wiggles in the redshift evolution of the evolution biases are due to the reconstruction of these functions from $n^{\rm evol}, s_\B$ and $s_\F$.

\revedit{To better understand which combination of splits provides the best constraints, we have performed a joint analysis combining the $50\times 50$ split with various other splits, ranging from 40\% of bright galaxies, down to 10\% of bright galaxies. The results are plotted in Fig.~\ref{fig:error_widths} of Appendix~\ref{app:tables}. We see that the more distinct the two samples are, the more the joint analysis improves the constraints with respect to the $50\times 50$ split. This is due to the fact that when the two splits are more distinct, the shot-noise covariance between them is smaller (see Appendix~\ref{sec:shotnoise}), which increases the amount of information gained by adding another split. As such, even if the most asymmetric case $10\times 90$ on its own does not always provide the best constraints (see left panel of Fig.~\ref{fig:error_widths}), it is the optimal one to combine with the $50\times 50$ case. Note however that for a large range of redshift the improvement saturates when decreasing the fraction of bright galaxies, and the combination with the $30\times 70$ case is almost as good as that with the $10\times 90$ case. }

Finally, in Table~\ref{tab:joint_cosmic} we show the constraints on the cosmological parameters. We find that these parameters are almost purely constrained by the even multipoles. Adding the odd multipoles only improves the constraints by 0.1\% at most. Note that this is consistent with the results of~\cite{Lorenz:2017iez}, who showed that relativistic effects are not useful to constrain cosmological parameters. Their capacity lies instead in constraining theories beyond $\Lambda$CDM~\cite{Tutusaus:2022cab,Castello:2023zjr,Bonvin:2018ckp,Bonvin:2020cxp,Castello:2022uuu,Bonvin:2022tii,Castello:2024jmq}, or magnification biases and evolution biases, as we show here. 

We compare the constraints on cosmological parameters using a single split and using the joint analysis. First we see that the constraints from the $30\times70$ split are always tighter than those from the $50\times 50$ split. This suggests that multi-tracer analyses should carefully explore various splits and determine the optimal one. The $50\times 50$ split is clearly the one which minimises shot noise in the cross-correlation signals (since one can form more pairs in this case), but since shot noise will be small in future surveys due to the very high number density of detected galaxies, this is not necessarily the optimal choice. The signal may indeed increase in other choices of splits, improving cosmological constraints. Note that these conclusions rely on our modelling of the galaxy biases, and may differ for a different modelling. Finally, we see that the joint analysis provides tighter constraints than the two single splits. The improvement with respect to the $30\times 70$ split lies between 1.5\% and 3\%. This is a somewhat limited improvement, but it comes at almost no cost. It does not require new data or different analysis techniques, but simply relies on doubling the size of the data vector by considering two types of cross-correlation.  

\clearpage

\section{Conclusions}\label{sec:conclusions}

In this work, we showed that exploiting the information carried by the odd multipoles of the 2-point correlation function can significantly improve our knowledge of the astrophysical properties of galaxies. We developed a physically motivated model for the magnification biases and evolution biases that govern the amplitude of the odd multipoles. These biases are directly related to the distribution in luminosity and the redshift evolution of galaxies and can therefore be used to reconstruct the luminosity function of galaxies. We found that:
\begin{itemize}
    \item The magnification biases and evolution biases of the various samples can all be related to three observable quantities: the magnification bias of the whole (un-split) population $s_\M$, the magnification bias of the bright sample $s_\B$, and the redshift evolution of the number of galaxies $\nevol{}$.
    \item These three quantities cannot be measured from the odd multipoles if we have no prior knowledge about them, since they are fully degenerate with each other. However, parametrising their redshift evolution and assuming a prior of 50\% on the parameters, the odd multipoles provide constraints that are tighter than the prior by a factor of up to $2-3$.
    \item Combining the dipole and the octupole is crucial to obtain tight constraints, since they do not suffer from the same degeneracies between parameters.
    \item Splitting the galaxies in different ways provides different constraints, but there is no significant trend that would single out one split as optimal.
    \item Combining two different splits further improves the constraints by up to 20\% on the biases and by $\sim 3\%$ on the cosmological parameters $n_{\rm s}$ and $\Omega_{\rm m}$.
\end{itemize}

Knowing the astrophysical properties of galaxies is necessary to robustly constrain local primordial non-Gaussianities from large-scale structure, since these are contaminated by relativistic effects at very large scales. Since these effects directly depend on the magnification and evolution biases, having precise measurements of these quantities is essential. Interestingly, one can use the odd multipoles to constrain these quantities at intermediate scales, where local primordial non-Gaussianities are irrelevant, and from this one can predict the contamination from relativistic effects to the even multipoles at very large scales, where primordial non-Gaussianities are measured. 

Finally, a good knowledge of the magnification and evolution biases is also essential to infer the observer velocity from the kinematic dipole, since they enter as a pre-factor in front of the observer velocity and are therefore fully degenerate with it. 
An accurate theoretical model of the kinematic dipole is critical for robust tests of the Cosmological Principle.

\acknowledgments 

C.B. and D.S.B. acknowledge support from the Swiss National Science Foundation. C.B. also acknowledges support from the European Research Council (ERC) under the European Union’s Horizon 2020 research and innovation program (grant agreement No. 863929; project title “Testing the law of gravity with novel large-scale structure observables”). 
R.M. is supported by the South African Radio Astronomy Observatory and the National Research Foundation (grant no. 75415).

\FloatBarrier

\appendix

\section{Modeling the shot-noise for overlapping populations}\label{sec:shotnoise}

\begin{figure}[h]
  \centering
  \begin{tikzpicture}
    \draw[line width=1.5pt] (0,0) rectangle (6,2);
    \draw[line width=1.5pt] (3,0) -- (3,2);
    
    \node at (0.5,1.5) {\textbf{B}};
    \node at (5.5,1.5) {\textbf{F}};
    
    \draw[line width=1.5pt] (0,-2.25) rectangle (6,-0.25);
    \draw[dashed] (3.0,-2.25) -- (3.0,-0.25);
    \draw[line width=1.5pt] (2.0,-2.25) -- (2.0,-0.25);

    \node at (0.5,-0.85) {\textbf{b}};
    \node at (5.5,-0.85) {\textbf{f}};
  \end{tikzpicture}
  \caption{Schematic representation of the overlap between two different splits. In the $m$ split we have more bright galaxies than in the $m'$ split, implying that some of the bright galaxies in the former are actually faint galaxies in the second case.
  }
  \label{fig:boxes}
\end{figure}
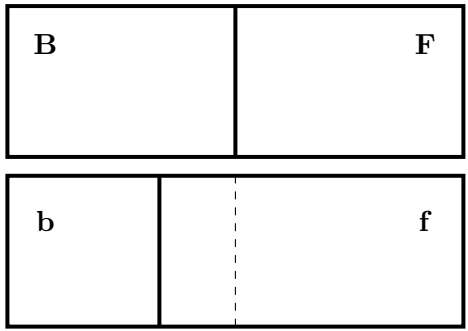

In this Appendix, we introduce a model describing the shot-noise contributions due to the overlap between two populations of galaxies. One usually assumes that, once the population is split into B and F galaxies, the populations are independent and therefore there is no shot noise contribution in the cross-correlation. This is appropriate for a single split, but it is inconsistent when combining two splits with two different values $m$ and $m'$ into the same analysis pipeline. In this case, there is automatically an overlap between the populations of the two different splits, as depicted in Fig.~\ref{fig:boxes}. This leads to a non-zero shot noise contribution between the splits.

Let us denote by B and F the bright and faint samples of the first split $m$,  and by b and f those of the second split $m'$. From Fig.~\ref{fig:boxes} we see that some of the B galaxies are the same as some of the f galaxies. Moreover, all $\bb$ galaxies are included in the B sample and all F galaxies are included in the $\f$ sample. Mathematically speaking, we can generally define the number density of each population of the $m'$ split as
\begin{align}
    & \bar{n}_\mathrm{b} = \theta_{\mathrm{Bb}}\,\bar{n}_\mathrm{B} + \theta_{\mathrm{Fb}}\,\bar{n}_\mathrm{F}, \\
    & \bar{n}_\mathrm{f} = \theta_{\mathrm{Bf}}\,\bar{n}_\mathrm{B} + \theta_{\mathrm{Ff}}\,\bar{n}_\mathrm{F},
\end{align}
where the $\theta$ parameters are numbers between $0$ and $1$ quantifying the overlap between the populations of each split. We need to find the specific values for each $m$ and $m'$ sub-samples combinations. In particular, for the case of a joint analysis combining the $50\times 50$ split with the $30\times 70$ split, we find:
\begin{equation}
    \theta_{\mathrm{\B\bb}} = \frac{3}{5},\quad \theta_{\mathrm{\F\bb}} = 0, \quad \theta_{\mathrm{\B\f}} = \frac{2}{5}, \quad \theta_{\mathrm{\F\f}} = 1.
\end{equation}
We immediately note that there is no overlap between the b and the F populations. 

In general, the fluctuations in the galaxy number count cross-correlations between the populations of each split have the form
\begin{equation}
    \langle\Delta_\LL(x_i)\,\Delta_\lambda(x_j)\rangle = C_{ij}^{\LL\lambda} + \frac{\theta_{\LL\lambda}}{\bar{n}_\lambda}\,\delta_{ij},
\end{equation}
where $L$, $\lambda$ label the galaxies luminosities in each split and $i$, $j$ label the pixel sky positions. We have denoted by $\Bar{n}_\lambda$ the background number density. The first term represents the cosmic variance contribution, which is non-zero even if the populations do not overlap, since they trace the same underlying matter density field. The second term encodes the shot-noise contribution, which depends directly on the overlapping faction $\theta_{\mathrm{L}\lambda}$: the less the populations overlap with each other, the smaller the shot-noise contribution is.

Considering two multipoles of orders $\ell$ and $\ell'$, the $m\times m'$ cross-variances will have the form 
\begin{equation}
    \mathrm{COV}\left(\xi^{(\ell)}_{\mathrm{LK}}(z,d),\,\xi^{(\ell')}_{\lambda\kappa}(z,d')\right) = \mathrm{COV}_\mathrm{CC} + \mathrm{COV}_\mathrm{CP} + \mathrm{COV}_\mathrm{P}\, ,
\end{equation}
where L,K $\in \{\B,\F\}$ and $\lambda, \kappa \in \{{\rm b}, {\rm f}\}$.
The last term contains the pure shot-noise contributions. For the dipole and octupole, the only non-vanishing contribution is when ${\rm L,K} =\B,\F$ and $\lambda, \kappa= {\rm b, f}$ (in all other cases, the signal itself vanishes) and it reads
\begin{align}
\mathrm{COV}_\mathrm{P}\left(\xi^{(\ell)}_{\B\F}(z,d),\,\xi^{(\ell')}_{\bb\f}(z,d')\right) = \frac{2\ell+1}{4\pi\,V\bar{N}^2_{\rm tot}l_{\rm p}}\frac{\delta_{d\,d'}\,\delta_{\ell\ell'}}{d^2}\frac{\theta_{\B\bb}\,\theta_{\F\f}}{q_\bb\,q_\f}\, \quad\mbox{for}\quad\ell,\ell'=1,3\, .
\end{align}
For the even multipoles however, there are also other non-vanishing correlations and one obtains
\begin{align}
    &\mathrm{COV}_\mathrm{P}\left(\xi^{(\ell)}_{\mathrm{LK}}(z,d),\,\xi^{(\ell')}_{\lambda\kappa}(z,d')\right) = \frac{2\ell+1}{4\pi\,V\bar{N}^2_{\rm tot}l_{\rm p}}\frac{\delta_{d\,d'}\,\delta_{\ell\ell'}}{d^2}\frac{1}{q_\lambda\,q_\kappa}\Big(\theta_{\LL\lambda}\,\theta_{\K\kappa}+\theta_{\LL\kappa}\,\theta_{\K\lambda}\Big)\,\label{eq:varp_1}\\
    &\quad\mbox{for}\quad\ell,\ell'=0,2,4\quad \mbox{and}\quad 
    {\rm L,K} \in \{\B,\F\}, \lambda, \kappa \in \{{\rm b}, {\rm f}\}\, .\nonumber
\end{align}
Here $V$ is the total volume of the survey, $\Bar{N}_{\rm tot}$ is the background total number density of galaxies, $l_{\rm p}$ is the pixel size and $q_\lambda=\bar{N}_\lambda/\bar{N}_g$ the fraction of galaxies of each type. In addition, there are also non-zero mixed (CP) contributions:
\begin{align}
    &\mathrm{COV}_\mathrm{CP}(\xi^{(\ell)}_{\mathrm{LK}}(z,d),\,\xi^{(\ell')}_{\lambda\kappa}(z,d')) = \left(\frac{D_1(z)}{D_1(0)}\right)^2\frac{1}{V \Bar{N}_{\rm tot}}\,(-1)^{\frac{\ell-\ell'}{2}}(-1)^{\,\ell+\ell'} \nonumber \\
    &\quad\times\frac{1}{8}\sum_a\Bigg[\frac{\theta_{\K\kappa}}{q_\kappa}\,\alpha_a^{\LL\lambda} + (-1)^{\ell+\ell'}\frac{\theta_{\LL\lambda}}{q_\lambda}\,\alpha_a^{\K\kappa}+(-1)^{\ell'}\left(\frac{\theta_{\K\lambda}}{q_\lambda}\,\alpha_a^{\LL\kappa} + (-1)^{\ell+\ell'}\frac{\theta_{\LL\kappa}}{q_\kappa}\,\alpha_a^{\K\lambda}\right)\Bigg] \nonumber \\
    &\quad\times \mathcal{G}(\ell,\ell',a)\,\mathcal{I}(\ell, \ell', d, d')\,, \label{eq:varcp}
\end{align}
where $D_1(z)$ is the growth function at redshift $z$ and the index $a$ runs over the values $0$, $2$, $4$. We have defined a set of three functions of the growth rate and the galaxy biases, denoted by $\alpha_a^{\mathrm{LK}}$. They take the form
\begin{align}
    &\alpha_0^{\mathrm{LK}} = b_\LL\,b_\K + \frac{f}{3}\,(b_\LL + b_\K) + \frac{f^2}{5}\,, \label{eq:a0}\\
    &\alpha_2^{\mathrm{LK}} = \frac{2}{3}\,(b_\LL + b_\K)\,f + \frac{4}{7}\,f^2\,,  \label{eq:a2}\\
    &\alpha_4^{\mathrm{LK}} = \frac{8}{35}\,f^2\,.  \label{eq:a4}
\end{align}
In addition, we also have 
\begin{equation}
    \mathcal{G}(\ell,\ell',a) = \int^{1}_{-1} d\mu\,P_\ell(\mu)\,P_{\ell'}(\mu)\,P_a(\mu)\,, 
    \label{eq:G} 
\end{equation}
where $\mu = \cos{\beta}$ is the angle formed by the direction of the incoming photons and the orientation of a pair of galaxies. Finally, we have the Fourier-Bessel transformations of the power spectrum at $z=0$
\begin{align}
    &\mathcal{I}(\ell,\ell',d,d') = \frac{2(2\ell+1)(2\ell'+1)}{\pi^2}\int\;dk\,k^2\;P_{\delta\delta}(k,z=0)\,j_\ell(kd)\;j_{\ell'}(kd')\,.\label{eq:I_even} 
\end{align}
Finally, in order to compute the CC contributions, we can use the usual expressions, replacing the corresponding values for the galaxy biases of each $m$, $m'$ sample~\cite{Bonvin:2015kuc,Hall:2016bmm}. The pure relativistic contributions to the cosmic variance relevant to the dipole and octupole covariances have been computed in \cite{Bonvin:2023jjq} and are also taken into account in this work.

\section{Covariance of the total hexadecapole with the monopole and quadrupole of the bright and faint populations}
\label{app:hexa}

The hexadecapole of the $2$-point correlation function is independent of the galaxy bias (see Eq.~\eqref{eq:hexa}) and it is therefore the same for each of the populations. As a consequence we only consider the hexadecapole of the whole population in our analysis. Since the whole population overlaps with the B and F populations, there is a non-zero shot noise contribution in the covariance of the hexadecapole with the monopole and quadrupole of the B and F populations. This can be computed using the formalism described in Appendix~\ref{app:hexa}, where we cross-correlate three populations instead of four (as in Fig.~\ref{fig:boxes}). In this case, we have
\begin{equation}
    \bar{n}_\mathrm{T} = \theta_{\mathrm{TB}}\,\bar{n}_{\mathrm{B}} + \theta_{\mathrm{TF}}\,\bar{n}_{\mathrm{F}},
\end{equation}
where $\theta_{\mathrm{TB}}=\theta_{\mathrm{TF}}=1$. The expressions are formally the same as Eqs.~\eqref{eq:varp_1}--\eqref{eq:varcp}. We use $b_\B$, $b_\F$ for the monopoles and quadrupoles ($\ell=0,\,2$) and $b_\mathrm{T}$ for the hexadecapole ($\ell'=4$). In this case the P and CP contributions are only sensitive to the total number density of galaxies, since the overlap of the B and F populations with the full population is total.

\section{Additional figures and tables}\label{app:tables}

In this Appendix, we provide additional figures and tables associated with the discussion and results developed in this work.

In Table~\ref{tab:SNRs_multipoles_} we show the cumulative SNR over all redshifts and separations for each of the multipoles and three different values of the sample separation parameter, $m=10/3,\,2.0,\,10/7$.

Tables~\ref{tab:fiducials_sB} and \ref{tab:fiducials_common} contain the fiducial values for all free parameters considered in the analyses. In Table~\ref{tab:fiducials_sB} we show the parameters that differ for each splits, whereas in Table~\ref{tab:fiducials_common} we show the parameters that do not depend on the split.

In Table~\ref{tab:constraints_magbias_prior20} we show the constraints on the various parameters starting with a $20\%$ prior, for three different splits. 

\begin{figure}[t]
    \centering
    \includegraphics[width=0.49\textwidth]{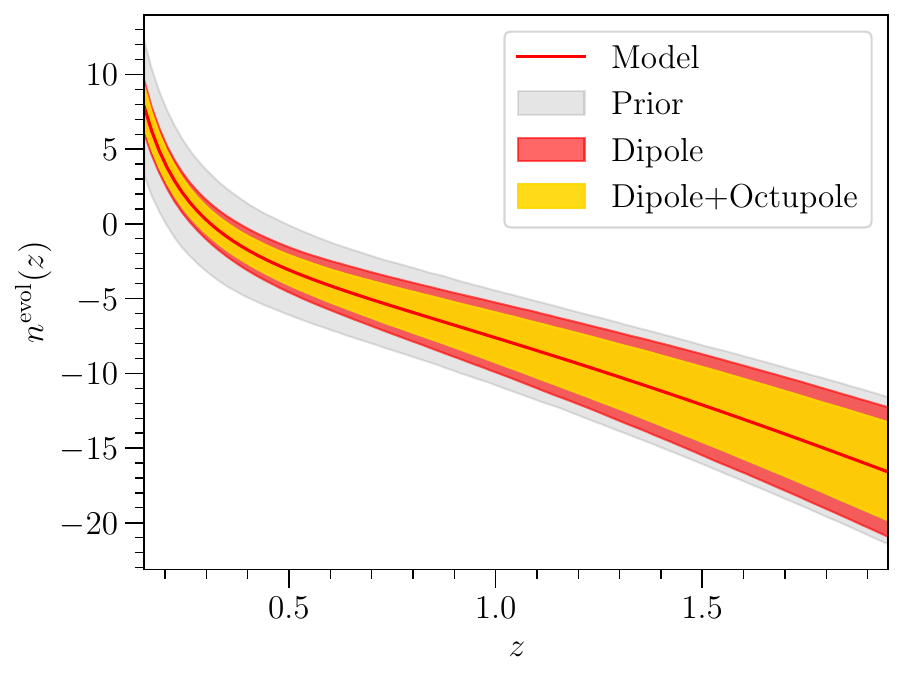}
    \includegraphics[width=0.48\textwidth]{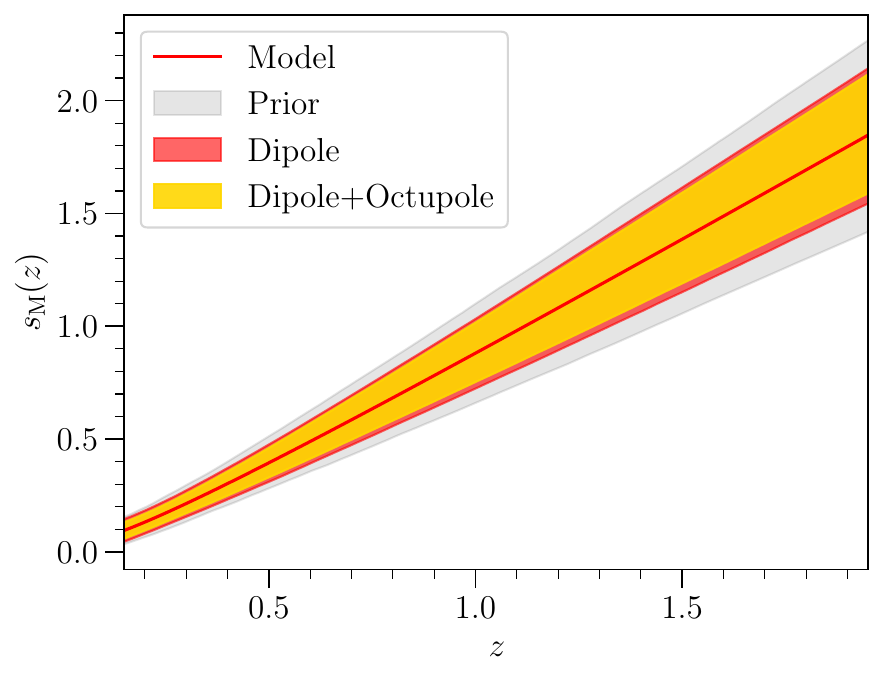}
    \includegraphics[width=0.48\textwidth]{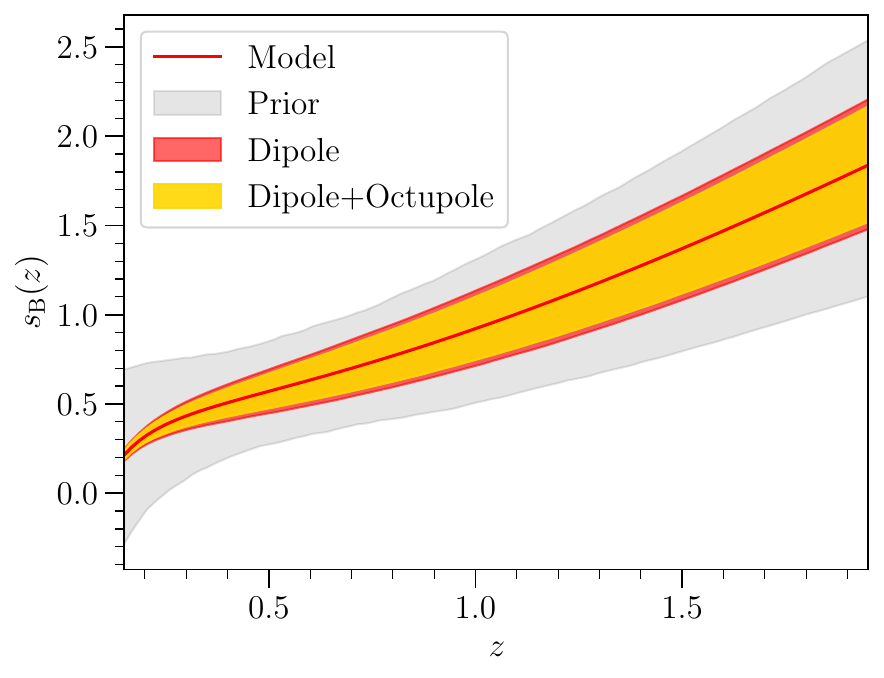}
    \includegraphics[width=0.49\textwidth]{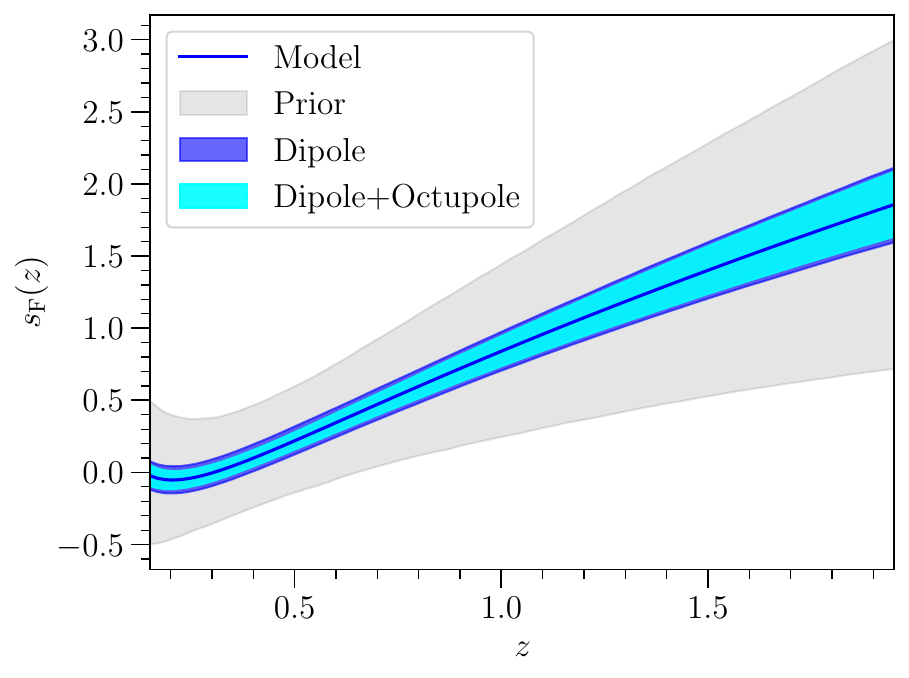}
    \includegraphics[width=0.49\textwidth]{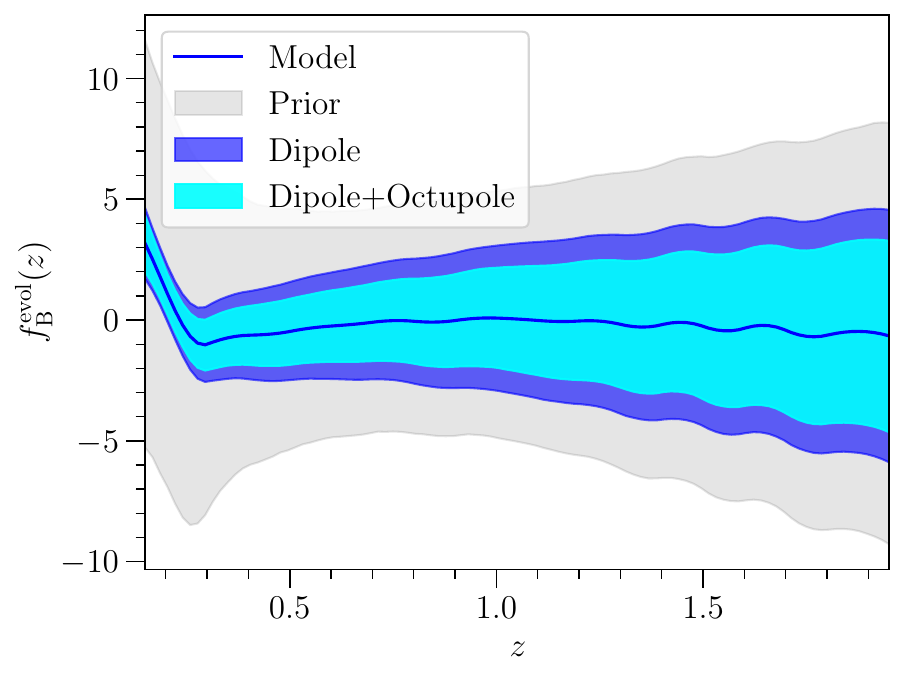}
    \includegraphics[width=0.49\textwidth]{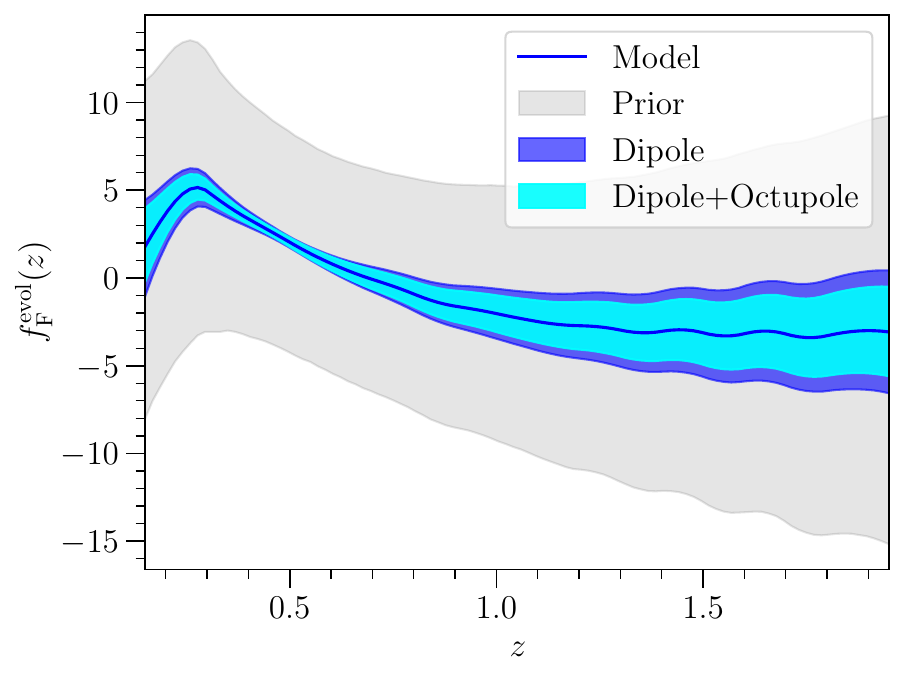}
    \caption{Constraints on the redshift evolution of the various functions for the 50$\times$50 split, starting with a 20\% prior on the parameters (grey region). We show the results when only the dipole is used and when both the dipole and octupole are included. In all cases the even multipoles (monopole, quadrupole and hexadecapole) are included. The red functions are those entering the forecasts, while the blue ones are derived from the red.}
    \label{fig:split50_prior20}
\end{figure}

\begin{figure}[t]
    \centering
    \includegraphics[width=0.49\textwidth]{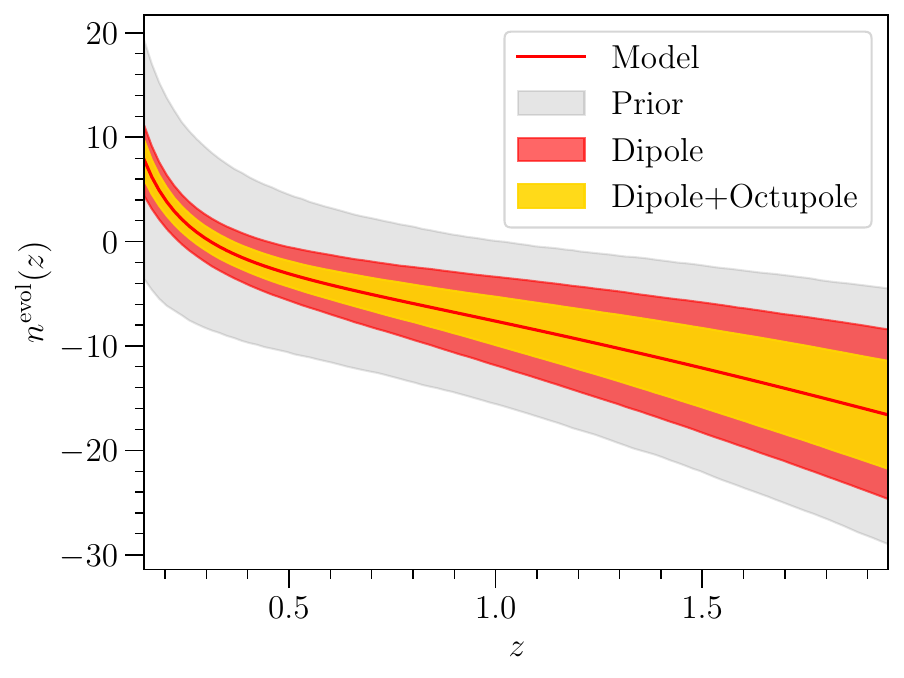}
    \includegraphics[width=0.48\textwidth]{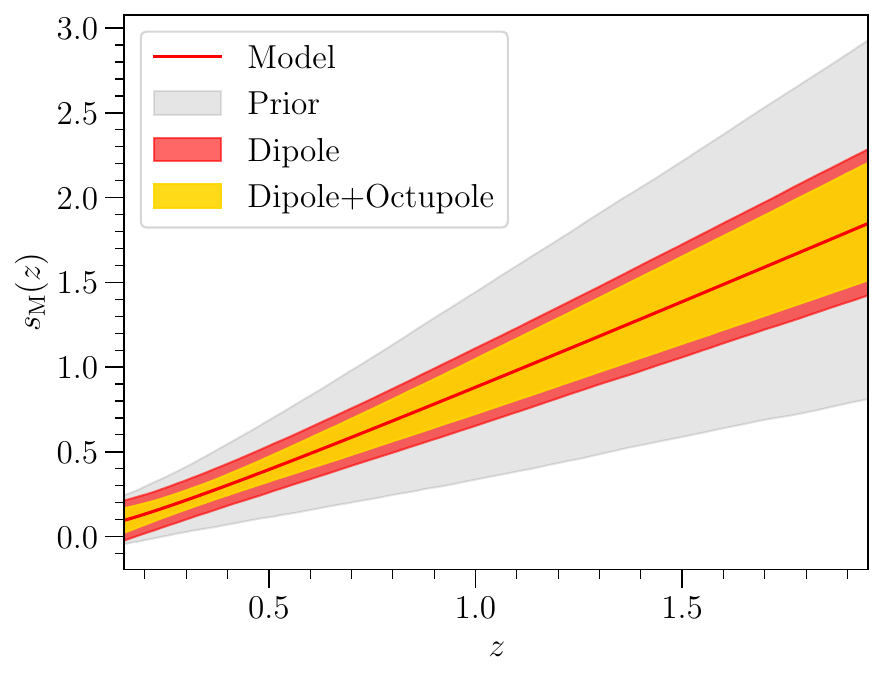}
    \includegraphics[width=0.49\textwidth]{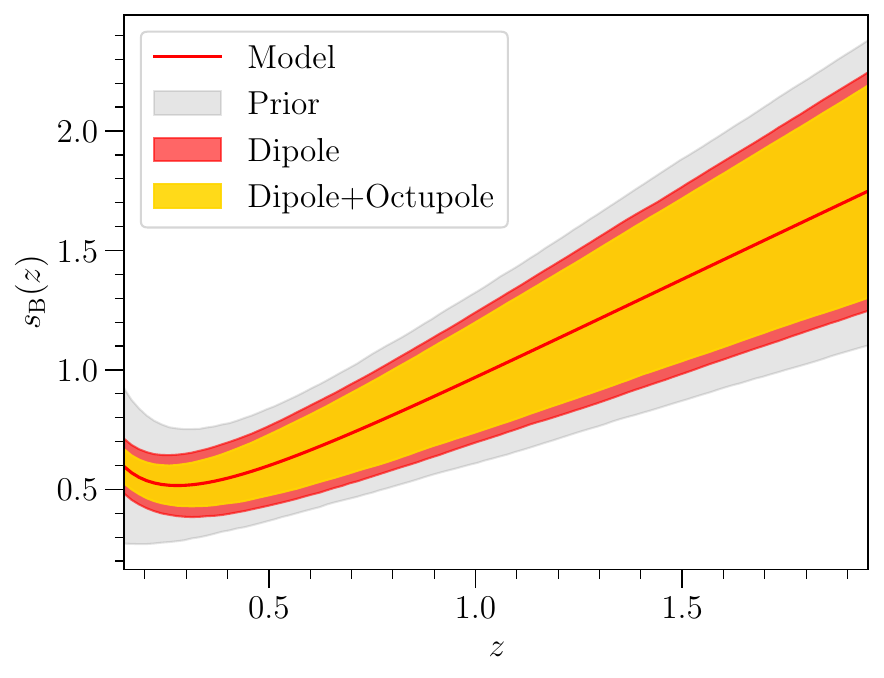}
    \includegraphics[width=0.475\textwidth]{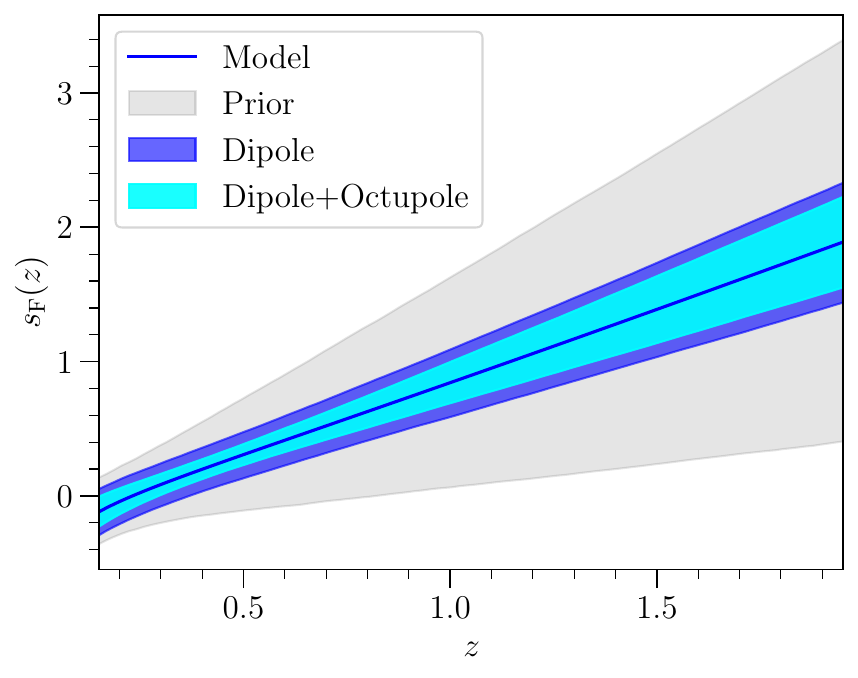}
    \includegraphics[width=0.49\textwidth]{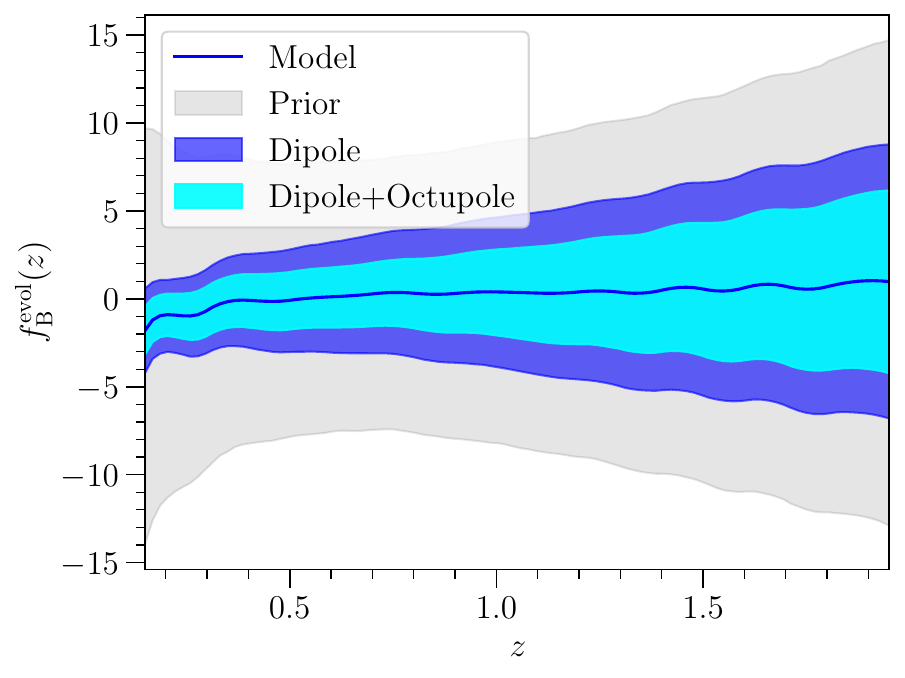}
    \includegraphics[width=0.49\textwidth]{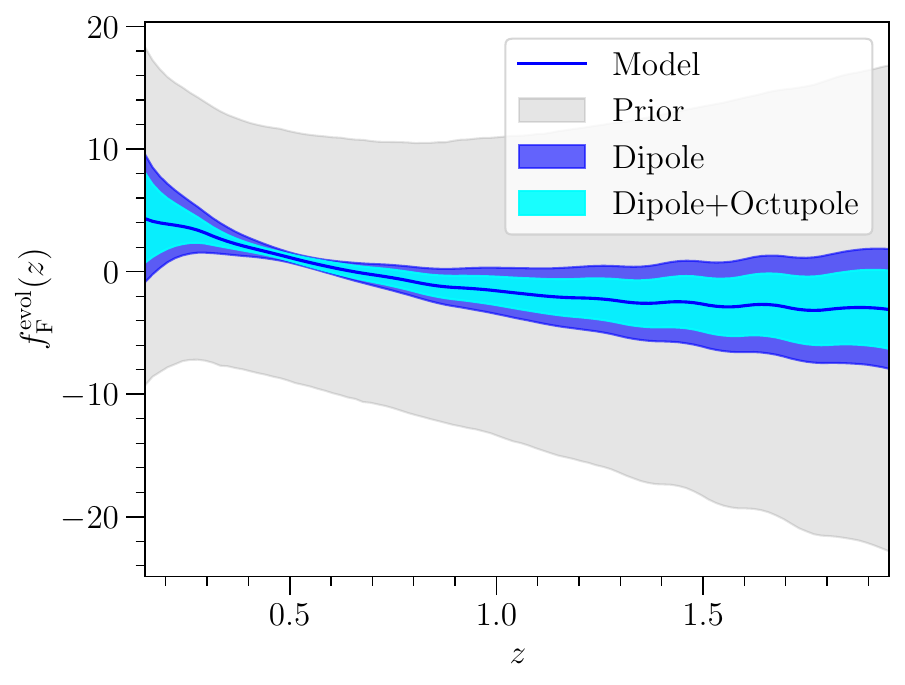}
    \caption{Constraints on the redshift evolution of the various functions for the 30$\times$70 split, starting with a 50\% prior on the parameters (grey region). We show the results when only the dipole is used and when both the dipole and octupole are included. In all cases the even multipoles (monopole, quadrupole and hexadecapole) are included. The red functions are those entering the forecasts, while the blue ones are derived from the red.}
    \label{fig:split30_prior50}
\end{figure}

\begin{figure}[t]
    \centering
    \includegraphics[width=\textwidth]{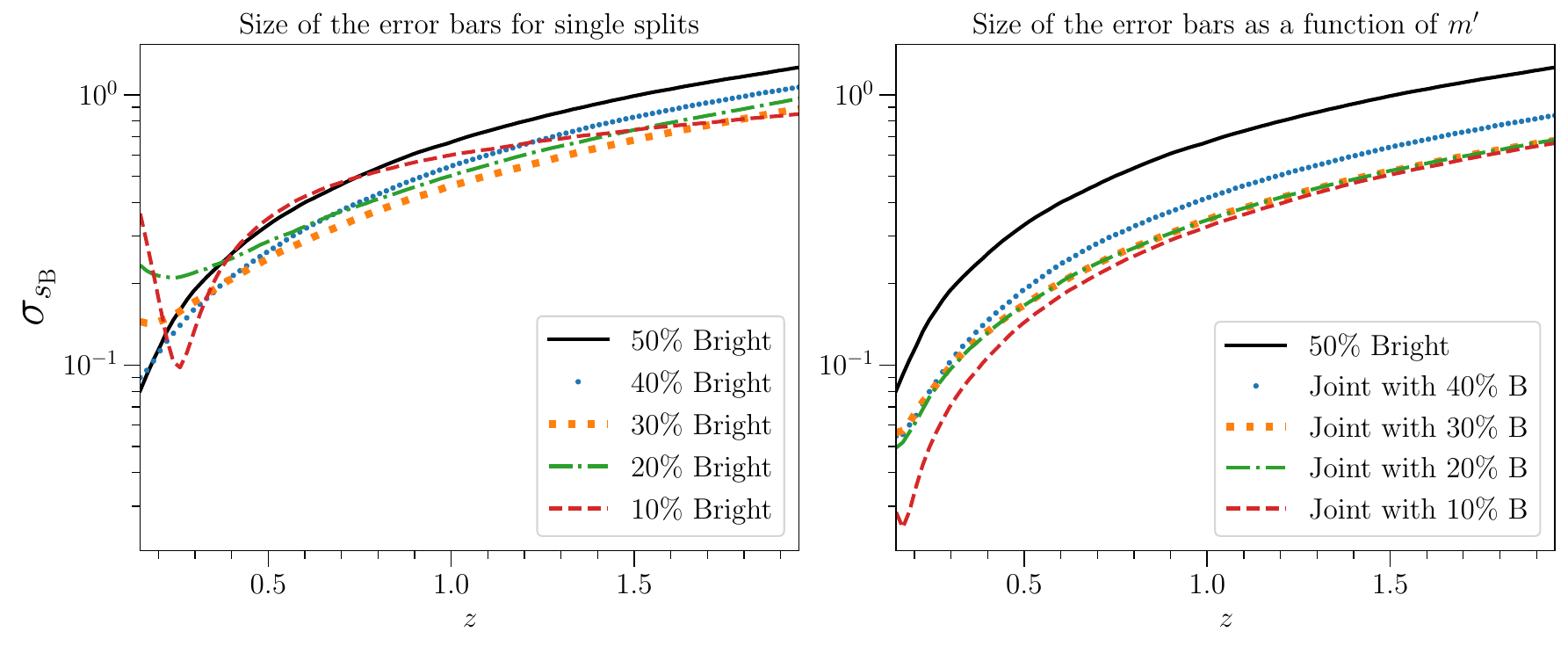}
    \includegraphics[width=\textwidth]{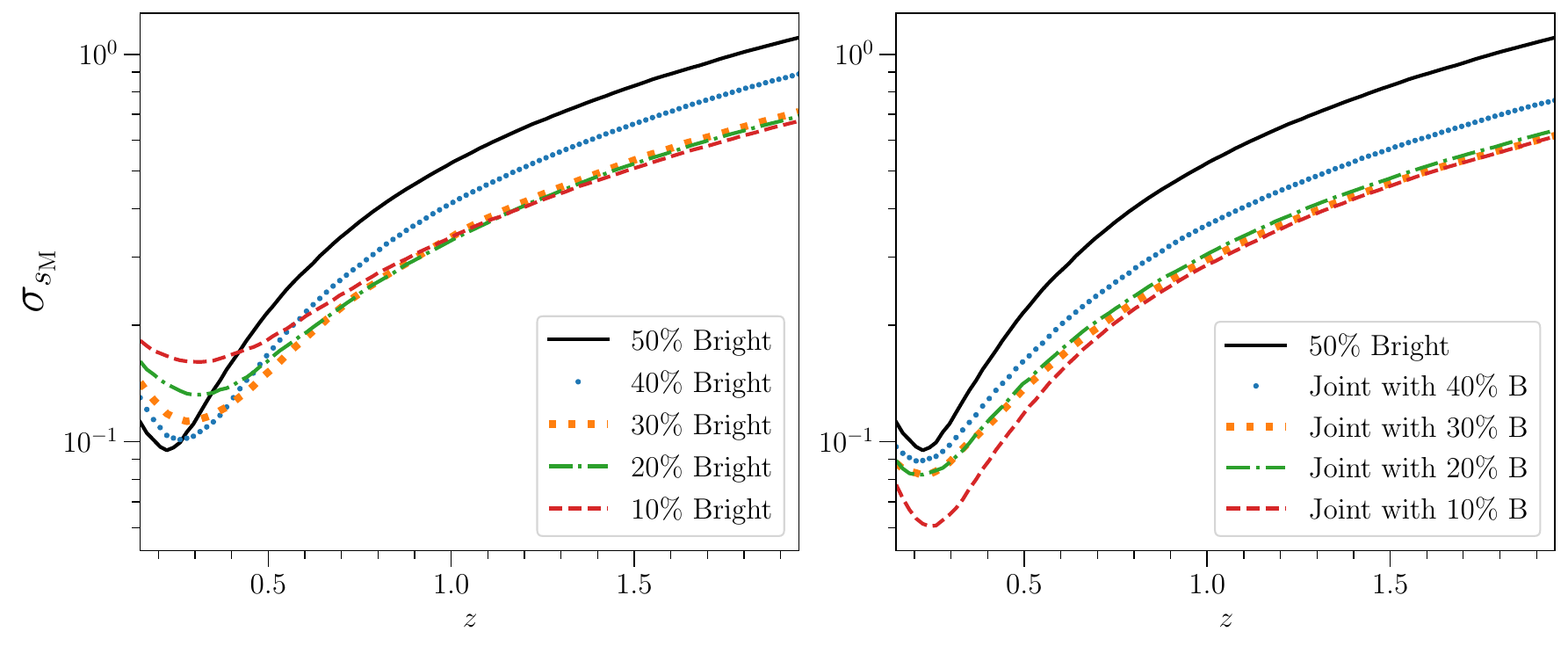}
    \includegraphics[width=\textwidth]{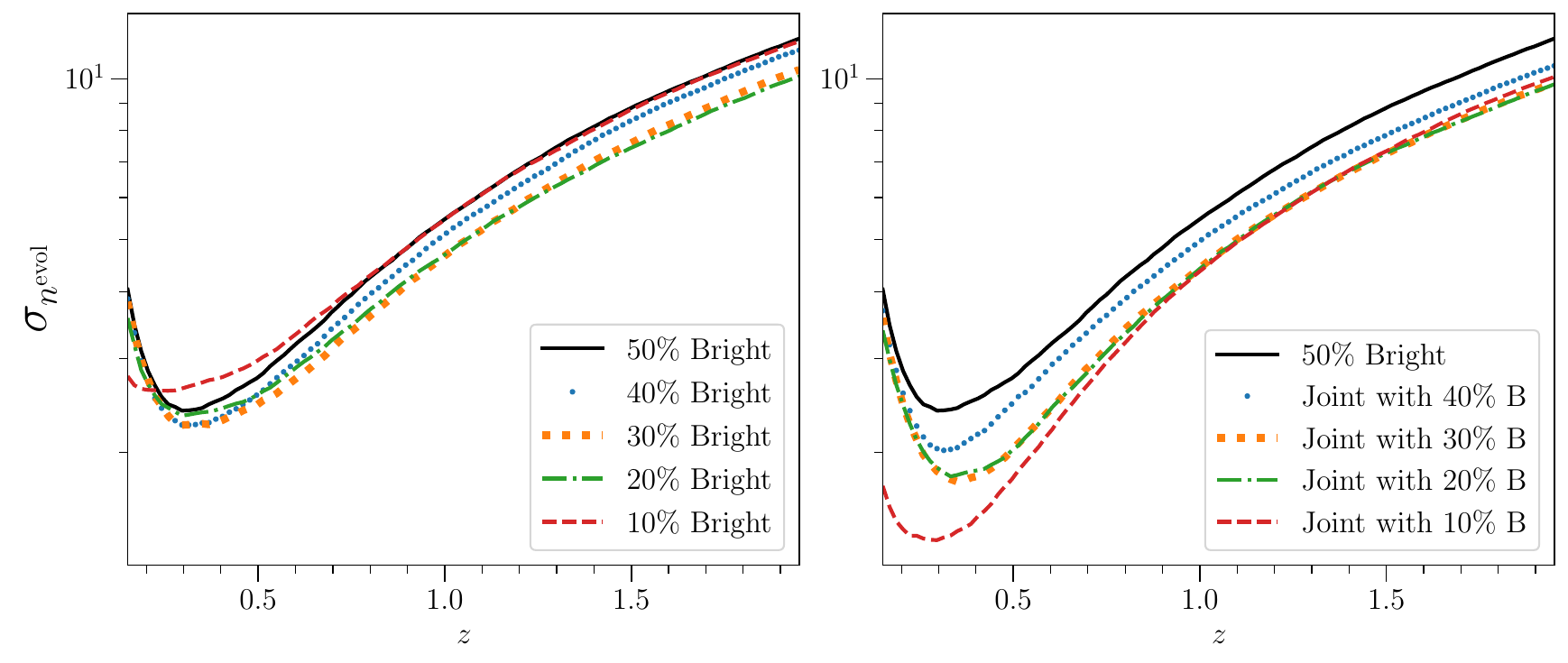}
    \caption{\revedit{Size of the error bars as a function of redshift for the parameters $s_\B$, $s_\M$ and $n_{\rm{evol}}$. The left panel shows the error bars for a single split, with different proportions of bright galaxies. The right panel compares the error bars for the $50\times 50$ split alone, with those of a joint analysis, combining the $50\times 50$ split with another of the other splits. }}
    \label{fig:error_widths}
\end{figure}

Finally, we include additional plots showing the redshift evolution of the parameters ($s_\B$, $s_\M$, $\nevol{}$) and the inferred quantities ($s_\F$, $\fevol{\B}$, $\fevol{\F}$), together with the predicted $68\%$ confidence regions. In Fig.~\ref{fig:split50_prior20} we show the results for the $50\times 50$ split when assuming a $20\%$ prior. In Fig.~\ref{fig:split30_prior50} we show the predictions for the $30\times 70$ split when assuming a $50\%$ prior. \revedit{In Fig.~\ref{fig:error_widths}, we show the size of the uncertainties on the parameters ($s_\B$, $s_\M$, $\nevol{}$) as a function of redshift for different splits, ranging from 50\% to 10\% of bright galaxies (left panel). We see that there is no clear trend: for some parameters and some redshift ranges the most extreme split ($10\times 90$) gives the best constraints, while in other cases it gives the worst constraints. This is related to the balance between the amplitude of the dipole signal and the shot noise contribution that both vary with the choice of split. In the right panel of Fig.~\ref{fig:error_widths}, we compare the uncertainties for the $50\times 50$ split alone, with those of a joint analysis of the $50\times 50$ split with another of the split. As discussed in the main text, we see that the improvement is always larger for the $10\times 90$ split.}

\begin{table}[h]
    \centering
    \begin{tabular}{lrrrrrrrrr}
        \toprule
        {} &  $\xi^{(0)}_{\B\B}$ &  $\xi^{(0)}_{\B\F}$ &  $\xi^{(0)}_{\F\F}$ &  $\xi^{(1)}_{\B\F}$ &  $\xi^{(2)}_{\B\B}$ &  $\xi^{(2)}_{\B\F}$ &  $\xi^{(2)}_{\F\F}$  &  $\xi^{(3)}_{\B\F}$ &  $\xi^{(4)}_{\rm T}$ \\
        \midrule
        $30\times70$ &   457.07 &   474.80 &   422.31 &      62.32 &   313.42 &   479.35 &   546.15 &      7.33 &  \\
        $50\times50$ &   521.12 &   426.13 &   301.99 &      55.47 &   391.46 &   522.47 &   489.21  &         4.78 &    74.50\\
        $70\times30$ &   671.17 &   571.24 &   395.93 &      51.87 &   497.11 &   601.89 &   543.65 &          6.45 & \\
        \bottomrule
    \end{tabular}
    \caption{Cumulative SNR over all redshifts and separations for the multipoles considered in the analysis. We show the SNR for three different splits. For the hexadecapole we show only one value corresponding to the whole population.}
    \label{tab:SNRs_multipoles_}
\end{table}

\begin{table}[h]
   \centering
   \begin{tabular}{lrrr}
   \toprule
    & $30\times 70$ & $50\times 50$ & $70\times 30$\\
   \midrule\midrule
       $s_{\B,0}$ & 0.3252 & -0.9011 & -0.9337 \\
       $s_{\B,1}$& 0.6442 & 1.8230 & 1.8243 \\
       $s_{\B,2}$ & 0.1614 & -1.0188 & -0.9284 \\
       $s_{\B,3}$ & 0.1333 & -0.3034 & -0.2410\\
   \bottomrule
   \end{tabular}
   \caption{Fiducial values of the $s_\B(z)$ fitting parameters. These are generally different for different splits.}
   \label{tab:fiducials_sB}
\end{table}

\begin{table}[h]
    \centering
    \begin{tabular}{lr}
        \toprule
        $s_{\M,0}$ & -0.1938 \\
       $s_{\M,1}$ & 1.0747 \\
       $s_{\M,2}$ & -0.0781  \\
       $s_{\M,3}$ & -0.0056  \\
       \midrule\midrule
       $\nevol{0}$ & 7.9460  \\
       $\nevol{1}$ & -12.0375 \\
       $\nevol{2}$ & 2.2993  \\
       $\nevol{3}$ & -15.8389  \\
       \midrule\midrule
       $b_{\B,1}$ & 0.5540  \\
       $b_{\B,2}$ & 0.7830   \\
       $b_{\F,1}$ & 0.5540   \\
       $b_{\F,2}$ & 0.7830   \\
       \midrule\midrule
       $h$ & 0.6766   \\
       $\ln{(10^{10} A_{\rm s})}$ & 3.0204  \\
       $n_{\rm s}$ & 0.9665   \\
       $\Omega_{\rm b}$ & 0.0490   \\
       $\Omega_{\rm m}$ & 0.3111  \\
       \bottomrule
\end{tabular}
    \caption{Fiducial values for the parameters varied in the analysis that are common to all splits.}
    \label{tab:fiducials_common}
\end{table}

\begin{table}[h]
    \centering
    \begin{tabular}{lrr}
        \toprule
        {} & \multicolumn{2}{c}{$50\times 50$ split} \\
        \cmidrule(lr){2-3} 
        {} &  $\xi^{(1)}_{\B\F}$ &  $\xi^{(1)}_{\B\F}+\xi^{(3)}_{\B\F}$ \\
        \midrule
        $s_{\B,0}$ &   15.03 &     13.83 \\
        $s_{\B,1}$ &   10.36 &      9.69 \\
        $s_{\B,2}$ &   10.72 &     10.60 \\
        $s_{\B,3}$ &   11.98 &     11.20 \\
        \cmidrule(lr){1-3}
        $s_{\M,0}$ &   19.18 &     17.16 \\
        $s_{\M,1}$ &   14.39 &     13.16 \\
        $s_{\M,2}$ &   19.73 &     18.81 \\
        $s_{\M,3}$ &   19.99 &     19.99 \\
        \midrule
        $\nevol{0}$ &   18.03 &     17.17 \\
        $\nevol{1}$ &   18.42 &     15.00 \\
        $\nevol{2}$ &   16.48 &     14.42 \\
        $\nevol{3}$ &   14.31 &     13.92 \\
        \bottomrule
    \end{tabular} 
    \begin{tabular}{rrr}
        \toprule
        \multicolumn{2}{c}{$30\times 70$ split} \\
        \cmidrule(lr){1-2}
        $\xi^{(1)}_{\B\F}$ &  $\xi^{(1)}_{\B\F}+\xi^{(3)}_{\B\F}$ \\
        \midrule
        17.45 &     16.25 \\
        17.60 &     16.93 \\
        17.93 &     17.73 \\
        12.36 &     11.75 \\
        \cmidrule(lr){1-2}
        19.31 &     17.47 \\
         8.86 &      8.37 \\
        19.83 &     19.29 \\
        19.99 &     19.99 \\
        \midrule
        16.85 &     16.65 \\
        16.62 &     14.49 \\
        14.73 &     13.45 \\
        13.44 &     13.29 \\
        \bottomrule
    \end{tabular}
    \begin{tabular}{rrr}
        \toprule
        \multicolumn{2}{c}{$70\times 30$ split} \\
        \cmidrule(lr){1-2}
        $\xi^{(1)}_{\B\F}$ &  $\xi^{(1)}_{\B\F}+\xi^{(3)}_{\B\F}$ \\
        \midrule
        12.50 &     11.71 \\
         9.28 &      8.95 \\
        10.49 &     10.33 \\
        13.11 &     12.60 \\
        \cmidrule(lr){1-2}
        18.51 &     16.29 \\
        14.73 &     13.89 \\
        19.28 &     17.76 \\
        19.98 &     19.96 \\
        \midrule
        18.63 &     17.35 \\
        19.36 &     16.14 \\
        15.49 &     14.45 \\
        15.52 &     14.55 \\
        \bottomrule
    \end{tabular}
    \caption{Relative 1$\sigma$ uncertainties for the magnification biases and number evolution parameters, assuming $20\%$ Gaussian priors. We show the results for 3 different splits, including only the dipole ($\xi_1$) and including both the dipole and octupole ($\xi_1+\xi_3$). In all cases the even multipoles (monopole, quadrupole and hexadecapole) are included.}
    \label{tab:constraints_magbias_prior20}
\end{table}


\FloatBarrier


\bibliographystyle{JHEP}
\bibliography{refs}


\end{document}